\documentclass[longauth]{aa}

\usepackage{euclid}
\usepackage{physics}
\usepackage{graphicx}
\usepackage{natbib}
\usepackage{scalerel}
\usepackage{multirow}   % multirow entries in tables

\usepackage[table]{xcolor}

\usepackage{tikz}
\newcommand{\tikzxmark}{%
\tikz[scale=0.23] {
    \draw[line width=0.7,line cap=round] (0,0) to [bend left=6] (1,1);
    \draw[line width=0.7,line cap=round] (0.2,0.95) to [bend right=3] (0.8,0.05);
}}

\usepackage{txfonts}
\usepackage[pdfencoding=auto,psdextra]{hyperref}
\hypersetup{
    colorlinks=true,
    linkcolor=blue,
    filecolor=magenta,      
    urlcolor=blue,
    citecolor=blue
}
\makeatletter
\renewcommand*\aa@pageof{, page \thepage{} of \pageref*{LastPage}}
\makeatother

\usepackage[utf8]{inputenc}

\usepackage[switch, modulo]{lineno}
\begin{document}

%
% Put the title of your paper here
%
\title{\Euclid preparation}
\subtitle{XXXIX.  The effect of baryons on the halo mass function}

%%%% please do not edit the author list -- contact ECEB Bureau for changes
\newcommand{\orcid}[1]{} %% define as link to https://orcid.org/#1 if needed
\author{Euclid Collaboration: T.~Castro\orcid{0000-0002-6292-3228}$^{1,2,3,4}$\thanks{\email{tiago.batalha@inaf.it}}, S.~Borgani\orcid{0000-0001-6151-6439}$^{5,3,1,2}$, M.~Costanzi\orcid{0000-0001-8158-1449}$^{1,5}$, J.~Dakin\orcid{0000-0002-2915-0315}$^{6}$, K.~Dolag$^{7}$, A.~Fumagalli\orcid{0009-0004-0300-2535}$^{5,1,2,3}$, A.~Ragagnin\orcid{0000-0002-8106-2742}$^{8,3,9}$, A.~Saro\orcid{0000-0002-9288-862X}$^{5,3,1,2,4}$, A.~M.~C.~Le~Brun\orcid{0000-0002-0936-4594}$^{10}$, N.~Aghanim$^{11}$
% ...
}

%%%% please do not edit the affiliation list -- contact ECEB Bureau for changes
\institute{$^{1}$ INAF-Osservatorio Astronomico di Trieste, Via G. B. Tiepolo 11, 34143 Trieste, Italy\\
$^{2}$ INFN, Sezione di Trieste, Via Valerio 2, 34127 Trieste TS, Italy\\
$^{3}$ IFPU, Institute for Fundamental Physics of the Universe, via Beirut 2, 34151 Trieste, Italy\\
$^{4}$ ICSC - Centro Nazionale di Ricerca in High Performance Computing, Big Data e Quantum Computing, Via Magnanelli 2, Bologna, Italy\\
$^{5}$ Dipartimento di Fisica - Sezione di Astronomia, Universit\`a di Trieste, Via Tiepolo 11, 34131 Trieste, Italy\\
$^{6}$ Institute for Computational Science, University of Zurich, Winterthurerstrasse 190, 8057 Zurich, Switzerland\\
$^{7}$ Universit\"ats-Sternwarte M\"unchen, Fakult\"at f\"ur Physik, Ludwig-Maximilians-Universit\"at M\"unchen, Scheinerstrasse 1, 81679 M\"unchen, Germany\\
$^{8}$ INAF-Osservatorio di Astrofisica e Scienza dello Spazio di Bologna, Via Piero Gobetti 93/3, 40129 Bologna, Italy\\
$^{9}$ Dipartimento di Fisica e Astronomia "Augusto Righi" - Alma Mater Studiorum Universit\`a di Bologna, via Piero Gobetti 93/2, 40129 Bologna, Italy\\
$^{10}$ Laboratoire Univers et Th\'eorie, Observatoire de Paris, Universit\'e PSL, Universit\'e Paris Cit\'e, CNRS, 92190 Meudon, France\\
$^{11}$ Universit\'e Paris-Saclay, CNRS, Institut d'astrophysique spatiale, 91405, Orsay, France\\
$^{12}$ Institute of Cosmology and Gravitation, University of Portsmouth, Portsmouth PO1 3FX, UK\\
$^{13}$ INAF-Osservatorio Astronomico di Brera, Via Brera 28, 20122 Milano, Italy\\
$^{14}$ Dipartimento di Fisica e Astronomia, Universit\`a di Bologna, Via Gobetti 93/2, 40129 Bologna, Italy\\
$^{15}$ INFN-Sezione di Bologna, Viale Berti Pichat 6/2, 40127 Bologna, Italy\\
$^{16}$ Max Planck Institute for Extraterrestrial Physics, Giessenbachstr. 1, 85748 Garching, Germany\\
$^{17}$ INAF-Osservatorio Astrofisico di Torino, Via Osservatorio 20, 10025 Pino Torinese (TO), Italy\\
$^{18}$ Dipartimento di Fisica, Universit\`a di Genova, Via Dodecaneso 33, 16146, Genova, Italy\\
$^{19}$ INFN-Sezione di Genova, Via Dodecaneso 33, 16146, Genova, Italy\\
$^{20}$ Department of Physics "E. Pancini", University Federico II, Via Cinthia 6, 80126, Napoli, Italy\\
$^{21}$ INAF-Osservatorio Astronomico di Capodimonte, Via Moiariello 16, 80131 Napoli, Italy\\
$^{22}$ INFN section of Naples, Via Cinthia 6, 80126, Napoli, Italy\\
$^{23}$ Instituto de Astrof\'isica e Ci\^encias do Espa\c{c}o, Universidade do Porto, CAUP, Rua das Estrelas, PT4150-762 Porto, Portugal\\
$^{24}$ Dipartimento di Fisica, Universit\`a degli Studi di Torino, Via P. Giuria 1, 10125 Torino, Italy\\
$^{25}$ INFN-Sezione di Torino, Via P. Giuria 1, 10125 Torino, Italy\\
$^{26}$ INAF-IASF Milano, Via Alfonso Corti 12, 20133 Milano, Italy\\
$^{27}$ Institut de F\'{i}sica d'Altes Energies (IFAE), The Barcelona Institute of Science and Technology, Campus UAB, 08193 Bellaterra (Barcelona), Spain\\
$^{28}$ Port d'Informaci\'{o} Cient\'{i}fica, Campus UAB, C. Albareda s/n, 08193 Bellaterra (Barcelona), Spain\\
$^{29}$ Institute for Theoretical Particle Physics and Cosmology (TTK), RWTH Aachen University, 52056 Aachen, Germany\\
$^{30}$ INAF-Osservatorio Astronomico di Roma, Via Frascati 33, 00078 Monteporzio Catone, Italy\\
$^{31}$ Dipartimento di Fisica e Astronomia "Augusto Righi" - Alma Mater Studiorum Universit\`a di Bologna, Viale Berti Pichat 6/2, 40127 Bologna, Italy\\
$^{32}$ Institute for Astronomy, University of Edinburgh, Royal Observatory, Blackford Hill, Edinburgh EH9 3HJ, UK\\
$^{33}$ Jodrell Bank Centre for Astrophysics, Department of Physics and Astronomy, University of Manchester, Oxford Road, Manchester M13 9PL, UK\\
$^{34}$ European Space Agency/ESRIN, Largo Galileo Galilei 1, 00044 Frascati, Roma, Italy\\
$^{35}$ ESAC/ESA, Camino Bajo del Castillo, s/n., Urb. Villafranca del Castillo, 28692 Villanueva de la Ca\~nada, Madrid, Spain\\
$^{36}$ University of Lyon, Univ Claude Bernard Lyon 1, CNRS/IN2P3, IP2I Lyon, UMR 5822, 69622 Villeurbanne, France\\
$^{37}$ Institute of Physics, Laboratory of Astrophysics, Ecole Polytechnique F\'ed\'erale de Lausanne (EPFL), Observatoire de Sauverny, 1290 Versoix, Switzerland\\
$^{38}$ UCB Lyon 1, CNRS/IN2P3, IUF, IP2I Lyon, 4 rue Enrico Fermi, 69622 Villeurbanne, France\\
$^{39}$ Mullard Space Science Laboratory, University College London, Holmbury St Mary, Dorking, Surrey RH5 6NT, UK\\
$^{40}$ Departamento de F\'isica, Faculdade de Ci\^encias, Universidade de Lisboa, Edif\'icio C8, Campo Grande, PT1749-016 Lisboa, Portugal\\
$^{41}$ Instituto de Astrof\'isica e Ci\^encias do Espa\c{c}o, Faculdade de Ci\^encias, Universidade de Lisboa, Campo Grande, 1749-016 Lisboa, Portugal\\
$^{42}$ Department of Astronomy, University of Geneva, ch. d'Ecogia 16, 1290 Versoix, Switzerland\\
$^{43}$ INAF-Istituto di Astrofisica e Planetologia Spaziali, via del Fosso del Cavaliere, 100, 00100 Roma, Italy\\
$^{44}$ Department of Physics, Oxford University, Keble Road, Oxford OX1 3RH, UK\\
$^{45}$ Universit\'e Paris-Saclay, Universit\'e Paris Cit\'e, CEA, CNRS, AIM, 91191, Gif-sur-Yvette, France\\
$^{46}$ INAF-Osservatorio Astronomico di Padova, Via dell'Osservatorio 5, 35122 Padova, Italy\\
$^{47}$ University Observatory, Faculty of Physics, Ludwig-Maximilians-Universit{\"a}t, Scheinerstr. 1, 81679 Munich, Germany\\
$^{48}$ Institute of Theoretical Astrophysics, University of Oslo, P.O. Box 1029 Blindern, 0315 Oslo, Norway\\
$^{49}$ Jet Propulsion Laboratory, California Institute of Technology, 4800 Oak Grove Drive, Pasadena, CA, 91109, USA\\
$^{50}$ von Hoerner \& Sulger GmbH, Schlo{\ss}Platz 8, 68723 Schwetzingen, Germany\\
$^{51}$ Technical University of Denmark, Elektrovej 327, 2800 Kgs. Lyngby, Denmark\\
$^{52}$ Cosmic Dawn Center (DAWN), Denmark\\
$^{53}$ Max-Planck-Institut f\"ur Astronomie, K\"onigstuhl 17, 69117 Heidelberg, Germany\\
$^{54}$ Department of Physics and Helsinki Institute of Physics, Gustaf H\"allstr\"omin katu 2, 00014 University of Helsinki, Finland\\
$^{55}$ Aix-Marseille Universit\'e, CNRS/IN2P3, CPPM, Marseille, France\\
$^{56}$ AIM, CEA, CNRS, Universit\'{e} Paris-Saclay, Universit\'{e} de Paris, 91191 Gif-sur-Yvette, France\\
$^{57}$ Universit\'e de Gen\`eve, D\'epartement de Physique Th\'eorique and Centre for Astroparticle Physics, 24 quai Ernest-Ansermet, CH-1211 Gen\`eve 4, Switzerland\\
$^{58}$ Department of Physics, P.O. Box 64, 00014 University of Helsinki, Finland\\
$^{59}$ Helsinki Institute of Physics, Gustaf H{\"a}llstr{\"o}min katu 2, University of Helsinki, Helsinki, Finland\\
$^{60}$ NOVA optical infrared instrumentation group at ASTRON, Oude Hoogeveensedijk 4, 7991PD, Dwingeloo, The Netherlands\\
$^{61}$ Universit\"at Bonn, Argelander-Institut f\"ur Astronomie, Auf dem H\"ugel 71, 53121 Bonn, Germany\\
$^{62}$ Aix-Marseille Universit\'e, CNRS, CNES, LAM, Marseille, France\\
$^{63}$ Department of Physics, Institute for Computational Cosmology, Durham University, South Road, DH1 3LE, UK\\
$^{64}$ Universit\'e C\^{o}te d'Azur, Observatoire de la C\^{o}te d'Azur, CNRS, Laboratoire Lagrange, Bd de l'Observatoire, CS 34229, 06304 Nice cedex 4, France\\
$^{65}$ European Space Agency/ESTEC, Keplerlaan 1, 2201 AZ Noordwijk, The Netherlands\\
$^{66}$ Universit\'e Paris-Saclay, Universit\'e Paris Cit\'e, CEA, CNRS, Astrophysique, Instrumentation et Mod\'elisation Paris-Saclay, 91191 Gif-sur-Yvette, France\\
$^{67}$ Space Science Data Center, Italian Space Agency, via del Politecnico snc, 00133 Roma, Italy\\
$^{68}$ Centre National d'Etudes Spatiales -- Centre spatial de Toulouse, 18 avenue Edouard Belin, 31401 Toulouse Cedex 9, France\\
$^{69}$ Institute of Space Science, Str. Atomistilor, nr. 409 M\u{a}gurele, Ilfov, 077125, Romania\\
$^{70}$ Instituto de Astrof\'isica de Canarias, Calle V\'ia L\'actea s/n, 38204, San Crist\'obal de La Laguna, Tenerife, Spain\\
$^{71}$ Departamento de Astrof\'isica, Universidad de La Laguna, 38206, La Laguna, Tenerife, Spain\\
$^{72}$ Dipartimento di Fisica e Astronomia "G. Galilei", Universit\`a di Padova, Via Marzolo 8, 35131 Padova, Italy\\
$^{73}$ INFN-Padova, Via Marzolo 8, 35131 Padova, Italy\\
$^{74}$ Departamento de F\'isica, FCFM, Universidad de Chile, Blanco Encalada 2008, Santiago, Chile\\
$^{75}$ Universit\"at Innsbruck, Institut f\"ur Astro- und Teilchenphysik, Technikerstr. 25/8, 6020 Innsbruck, Austria\\
$^{76}$ Institut d'Estudis Espacials de Catalunya (IEEC), Carrer Gran Capit\'a 2-4, 08034 Barcelona, Spain\\
$^{77}$ Institute of Space Sciences (ICE, CSIC), Campus UAB, Carrer de Can Magrans, s/n, 08193 Barcelona, Spain\\
$^{78}$ Satlantis, University Science Park, Sede Bld 48940, Leioa-Bilbao, Spain\\
$^{79}$ Centro de Investigaciones Energ\'eticas, Medioambientales y Tecnol\'ogicas (CIEMAT), Avenida Complutense 40, 28040 Madrid, Spain\\
$^{80}$ Instituto de Astrof\'isica e Ci\^encias do Espa\c{c}o, Faculdade de Ci\^encias, Universidade de Lisboa, Tapada da Ajuda, 1349-018 Lisboa, Portugal\\
$^{81}$ Universidad Polit\'ecnica de Cartagena, Departamento de Electr\'onica y Tecnolog\'ia de Computadoras,  Plaza del Hospital 1, 30202 Cartagena, Spain\\
$^{82}$ Institut de Recherche en Astrophysique et Plan\'etologie (IRAP), Universit\'e de Toulouse, CNRS, UPS, CNES, 14 Av. Edouard Belin, 31400 Toulouse, France\\
$^{83}$ Kapteyn Astronomical Institute, University of Groningen, PO Box 800, 9700 AV Groningen, The Netherlands\\
$^{84}$ INFN-Bologna, Via Irnerio 46, 40126 Bologna, Italy\\
$^{85}$ Infrared Processing and Analysis Center, California Institute of Technology, Pasadena, CA 91125, USA\\
$^{86}$ Institut d'Astrophysique de Paris, UMR 7095, CNRS, and Sorbonne Universit\'e, 98 bis boulevard Arago, 75014 Paris, France\\
$^{87}$ CEA Saclay, DFR/IRFU, Service d'Astrophysique, Bat. 709, 91191 Gif-sur-Yvette, France\\
$^{88}$ Institut f\"ur Theoretische Physik, University of Heidelberg, Philosophenweg 16, 69120 Heidelberg, Germany\\
$^{89}$ Universit\'e St Joseph; Faculty of Sciences, Beirut, Lebanon\\
$^{90}$ Institut d'Astrophysique de Paris, 98bis Boulevard Arago, 75014, Paris, France\\
$^{91}$ Junia, EPA department, 41 Bd Vauban, 59800 Lille, France\\
$^{92}$ SISSA, International School for Advanced Studies, Via Bonomea 265, 34136 Trieste TS, Italy\\
$^{93}$ Instituto de F\'isica Te\'orica UAM-CSIC, Campus de Cantoblanco, 28049 Madrid, Spain\\
$^{94}$ CERCA/ISO, Department of Physics, Case Western Reserve University, 10900 Euclid Avenue, Cleveland, OH 44106, USA\\
$^{95}$ Dipartimento di Fisica e Scienze della Terra, Universit\`a degli Studi di Ferrara, Via Giuseppe Saragat 1, 44122 Ferrara, Italy\\
$^{96}$ Istituto Nazionale di Fisica Nucleare, Sezione di Ferrara, Via Giuseppe Saragat 1, 44122 Ferrara, Italy\\
$^{97}$ NASA Ames Research Center, Moffett Field, CA 94035, USA\\
$^{98}$ Kavli Institute for Particle Astrophysics \& Cosmology (KIPAC), Stanford University, Stanford, CA 94305, USA\\
$^{99}$ Bay Area Environmental Research Institute, Moffett Field, California 94035, USA\\
$^{100}$ Minnesota Institute for Astrophysics, University of Minnesota, 116 Church St SE, Minneapolis, MN 55455, USA\\
$^{101}$ INAF, Istituto di Radioastronomia, Via Piero Gobetti 101, 40129 Bologna, Italy\\
$^{102}$ Institute Lorentz, Leiden University, PO Box 9506, Leiden 2300 RA, The Netherlands\\
$^{103}$ Institute for Astronomy, University of Hawaii, 2680 Woodlawn Drive, Honolulu, HI 96822, USA\\
$^{104}$ Department of Physics \& Astronomy, University of California Irvine, Irvine CA 92697, USA\\
$^{105}$ Departamento F\'isica Aplicada, Universidad Polit\'ecnica de Cartagena, Campus Muralla del Mar, 30202 Cartagena, Murcia, Spain\\
$^{106}$ Department of Astronomy \& Physics and Institute for Computational Astrophysics, Saint Mary's University, 923 Robie Street, Halifax, Nova Scotia, B3H 3C3, Canada\\
$^{107}$ Universit\'e Paris Cit\'e, CNRS, Astroparticule et Cosmologie, 75013 Paris, France\\
$^{108}$ Department of Computer Science, Aalto University, PO Box 15400, Espoo, FI-00 076, Finland\\
$^{109}$ Ruhr University Bochum, Faculty of Physics and Astronomy, Astronomical Institute (AIRUB), German Centre for Cosmological Lensing (GCCL), 44780 Bochum, Germany\\
$^{110}$ Universit\'e Paris-Saclay, CNRS/IN2P3, IJCLab, 91405 Orsay, France\\
$^{111}$ Univ. Grenoble Alpes, CNRS, Grenoble INP, LPSC-IN2P3, 53, Avenue des Martyrs, 38000, Grenoble, France\\
$^{112}$ Department of Physics and Astronomy, Vesilinnantie 5, 20014 University of Turku, Finland\\
$^{113}$ Serco for European Space Agency (ESA), Camino bajo del Castillo, s/n, Urbanizacion Villafranca del Castillo, Villanueva de la Ca\~nada, 28692 Madrid, Spain\\
$^{114}$ Centre for Astrophysics \& Supercomputing, Swinburne University of Technology, Victoria 3122, Australia\\
$^{115}$ ARC Centre of Excellence for Dark Matter Particle Physics, Melbourne, Australia\\
$^{116}$ Oskar Klein Centre for Cosmoparticle Physics, Department of Physics, Stockholm University, Stockholm, SE-106 91, Sweden\\
$^{117}$ Astrophysics Group, Blackett Laboratory, Imperial College London, London SW7 2AZ, UK\\
$^{118}$ Centre de Calcul de l'IN2P3/CNRS, 21 avenue Pierre de Coubertin 69627 Villeurbanne Cedex, France\\
$^{119}$ Dipartimento di Fisica, Sapienza Universit\`a di Roma, Piazzale Aldo Moro 2, 00185 Roma, Italy\\
$^{120}$ INFN-Sezione di Roma, Piazzale Aldo Moro, 2 - c/o Dipartimento di Fisica, Edificio G. Marconi, 00185 Roma, Italy\\
$^{121}$ Centro de Astrof\'{\i}sica da Universidade do Porto, Rua das Estrelas, 4150-762 Porto, Portugal\\
$^{122}$ Zentrum f\"ur Astronomie, Universit\"at Heidelberg, Philosophenweg 12, 69120 Heidelberg, Germany\\
$^{123}$ Dipartimento di Fisica, Universit\`a di Roma Tor Vergata, Via della Ricerca Scientifica 1, Roma, Italy\\
$^{124}$ INFN, Sezione di Roma 2, Via della Ricerca Scientifica 1, Roma, Italy\\
$^{125}$ Leiden Observatory, Leiden University, Niels Bohrweg 2, 2333 CA Leiden, The Netherlands\\
$^{126}$ Department of Physics and Astronomy, University of California, Davis, CA 95616, USA\\
$^{127}$ Department of Astrophysical Sciences, Peyton Hall, Princeton University, Princeton, NJ 08544, USA\\
$^{128}$ Niels Bohr Institute, University of Copenhagen, Jagtvej 128, 2200 Copenhagen, Denmark\\
$^{129}$ Cosmic Dawn Center (DAWN)}

% 
% Put your abstract here:
%
\abstract{
The \Euclid photometric survey of galaxy clusters stands as a powerful cosmological tool, with the capacity to significantly propel our understanding of the Universe. Despite being subdominant to dark matter and dark energy, the baryonic component of our Universe holds substantial influence over the structure and mass of galaxy clusters. This paper presents a novel model that can be used to precisely quantify the impact of baryons on the  virial halo masses of galaxy clusters using the baryon fraction within a cluster as a proxy for their effect. Constructed on the premise of quasi-adiabaticity, the model includes two parameters, which are calibrated using non-radiative cosmological hydrodynamical simulations, and a single large-scale simulation from the Magneticum set, which includes the physical processes driving galaxy formation. As a main result of our analysis, we demonstrate that this model delivers a remarkable 1\%\ relative accuracy in determining the virial dark matter-only equivalent mass of galaxy clusters starting from the corresponding total cluster mass and baryon fraction measured in hydrodynamical simulations. Furthermore, we demonstrate that this result is robust against changes in cosmological parameters and against variation of the numerical implementation of the subresolution physical processes included in the simulations. Our work substantiates previous claims regarding the impact of baryons on cluster cosmology studies. In particular, we show how neglecting these effects would lead to biased cosmological constraints for a \Euclid-like cluster abundance analysis. Importantly, we demonstrate that uncertainties associated with our model arising from baryonic corrections to cluster masses are subdominant when compared to the precision with which mass--observable (i.e.\ richness) relations will be calibrated using \Euclid and to our current understanding of the baryon fraction within galaxy clusters.}
%
% Provide up to five key words:
%
%
% Add short versions of title and author list for page headings
%
   \titlerunning{\Euclid preparation. XXXIX. The effect of baryons on the HMF}
   \authorrunning{Castro et al.}
   
   \maketitle
%
%-------------------------------------------------------------------
%
%
%   Start the main text of your paper here
%
   
\section{\label{sec:intro}Introduction}
In the Lambda Cold Dark Matter ($\Lambda$CDM) scenario, structures are formed hierarchically, with larger systems emerging from the merger of smaller ones. As the largest, most massive, virialized objects in the Universe, galaxy clusters are at the pinnacle of this hierarchy and provide competitive cosmological probes of the geometry of our Universe and of the growth of density perturbations through measurements of their cosmic abundance and clustering~\citep[see e.g. ][]{Allen:2011zs, Kravtsov:2012zs, Fumagalli:2023yym}. 

While structure formation is gravitationally dominated by dark matter, astrophysical effects associated with the baryonic component are known to significantly affect the clusters and cluster galaxy population ~\citep[see e.g.][]{McDonald:2012hz,Webb:2015tha,2019A&A...628A..34E,Schellenberger:2019aws,2020NatAs...4..957Y,Debackere:2021ado}. In this context, advanced cosmological simulations~\citep[e.g.][]{Borgani:2009cd} provide the best tools for studying the formation of galaxy clusters starting from primordial density perturbations.

The primary cosmological probe from cluster surveys is derived from number-count experiments~\citep[][]{borgani:2001,holder:2001db, DSDD:2009php, Hasselfield:2013wf,Planck:2013lkt,SPT:2014wkb,Mantz:2014paa,Planck:2015lwi,SPT:2018njh, DES:2020ahh,DES:2020cbm, Lesci:2020qpk}, which rely on the strong dependence of both volumes and the density of halos as a function of mass and redshift on cosmological parameters. The number density of halos as a function of mass and redshift is called the halo mass function (HMF).

The prediction of the HMF relies deeply on the results of cosmological simulations~\citep[see e.g.][]{Tinker:2008ff,Watson:2012mt,Bocquet:2015pva,Despali:2015yla, Castro:2020yes,Bocquet:2020tes,Euclid:2022dbc}. As simulations involving full hydrodynamical calculations are more expensive than purely gravitational $N$-body simulations, and the cost can even be prohibitive, the conventional method is to characterise the HMF using the latter and model how baryonic physics alters the mass of halos in post-processing~\citep[][]{Schneider:2015wta,Arico:2020lhq}. For the sake of brevity, we refer to these hydrodynamical simulations simply as `hydro' and their dark-matter(DM)-only $N$-body counterparts as `dmo'. As mentioned, despite being subdominant, the baryonic component has a sizeable impact on the detailed properties of the large-scale structure of our Universe~\citep{Cui:2014aga,Velliscig:2014bza,Bocquet:2015pva,Castro:2017tbn,Castro:2020yes,10.1093/mnras/stad2419}.  

Feedback processes related to supernovae (SNe) and to active galactic nuclei (AGN) originate on small scales that cannot be explicitly resolved in simulations covering large cosmological volumes. As a result, it is impossible to explicitly simulate these processes within such large volumes from first principles; instead, subresolution models must be used~\citep[e.g.][]{Springel:2003,Hirschmann:2013qfl,Vogelsberger:2014dza,Schaye:2014tpa,Crain:2015poa,McCarthy:2016mry,Vogelsberger:2019ynw}. However, these processes affect the distribution of baryons and cosmic structures on scales well resolved by cosmological simulations and probed by observations~\citep[see e.g.][]{vanDaalen:2011xb}. 

The strategy of describing such processes in simulations through subresolution phenomenological models has offered invaluable insights for quantifying the influence of baryonic effects on large-scale structure~\citep[see e.g.][]{Teyssier:2010dp,vanDaalen:2011xb,Martizzi:2012ci,Sawala:2015cdf,Bullock:2017xww,vanDaalen:2019pst,10.1093/mnras/stad2419}. However, due to our incomplete understanding of the associated subresolution physical processes, several of the parameters used in their implementation require a calibration to reproduce some key observables. This calibration is usually carried out for one or a few specific observables, and is intrinsically resolution dependent~\citep[see][for discussion]{Schaye:2014tpa}. Furthermore, due to the complex interaction and the degeneracy between highly non-linear processes, different choices for such parameters can equally reproduce the target observables.

It is established that at the scales relevant for galaxy clusters, baryonic feedback cannot disrupt structures. However, various astrophysical processes associated with the baryonic component can still alter the halo mass. Specifically, these processes include radiative cooling and star formation, both of which can induce adiabatic contraction of the halo. Moreover, sudden displacement of gas caused by impulsive AGN feedback can lead to expansion of the halo. These dynamic processes significantly modify the halo mass when compared to the same halo described within a dmo simulation ~\citep[see e.g.][]{Velliscig:2014bza,Castro:2020yes}. In this light, a comprehensive understanding of such influences is indispensable in order to accurately model and interpret hydro simulations.

In the present paper, we model the impact of  baryonic effects on the halo virial mass. Our model assumes that the processes involved are quasi-adiabatic; that is, that they happen in a steady, non-abrupt manner. Although our model is calibrated on a set of non-radiative hydro simulations and on a single realization of a full-physics simulation, we demonstrate that its performance is robust against the change of parameters controlling the subresolution physics and our choice between independent implementations of subresolution processes. 

The large statistics of clusters that will be provided by the \Euclid~\citep[][]{2011arXiv1110.3193L} cluster survey require a high-precision calibration of the HMF, so that uncertainties on this theory-predicted quantity do not represent a limiting factor in the precision achievable by the cluster survey~\citep[see][]{Euclid:2019bue}. In \cite{Euclid:2022dbc}, we used a large set of $N$-body simulations to  derive  an analytical, cosmology-dependent expression for the HMF that meets the required precision of $\simeq 1$ \%. In the present paper, we extend the results of our previous work to account for baryonic effects. Our goal is therefore to reach a 1\%\ accuracy on the baryonic impact description in order to match the \Euclid requirements and allow optimum exploitation of the cosmological constraining power of the \Euclid cluster survey~\citep{Sartoris:2015aga}.

This paper is organized as follows: we outline the methodology used in this paper in Sect.~\ref{sec:methodology}. In Sect.~\ref{sec:results}, we present our findings from the simulations and the performance of the baryonic correction model. In Sect.~\ref{sec:euclid}, we assess the impact of our model in a forecast \Euclid cluster-count analysis. Some final remarks are provided in Sect.~\ref{sec:conclusions}. Lastly, in Appendix~\ref{app:content}, we study the baryonic content in different simulations.

\section{\label{sec:methodology}Methodology}
\subsection{\label{sec:sims}Simulations}
In this section, we describe the simulations used to calibrate the baryonic effects on halo masses, and the preparation of the halo catalogues used to compare results from hydro and dmo simulations. 

\subsubsection{The Magneticum set}
The Magneticum\footnote{\url{http://www.magneticum.org}} simulations were carried out using the TreePM+SPH code P-Gadget3, which is a more efficient variation of Gadget-2~\citep{Springel:2000yr,Springel:2005mi}. The SPH solver uses the revised implementation of~\citet{Beck:2015qva}, which overcomes a number of limitations of the traditional SPH solvers. The Magneticum hydro simulations include treatment of radiative cooling, heating by a uniform, evolving UV background, star formation and stellar feedback following \cite{Springel:2004kf}, and the description of stellar evolution and chemical enrichment processes by \cite{Tornatore:2007ds}. According to the latter, 11 chemical elements are followed, once produced by AGB stars, Type-Ia SN, and Type-II SN (H; He; C; N; O; Ne; Mg; Si; S; Ca; Fe). Following~\citet{Wiersma:2008cs}, metallicity-dependent cooling is implemented using cooling tables generated by the freely accessible CLOUDY photo-ionization algorithm~\citep{Ferland:1998id}. Supermassive black holes, which are hosted at the centre of galaxies, are described as sink particles whose mass increases by gas accretion and merging with other BHs ~\citep{Springel:2004kf,DiMatteo:2007sq}. Accretion onto these black holes is governed by the Bondi-Hoyle-Lyttleton formula~\citep{hoyle_lyttleton_1939,Bondi:1944jm,Bondi:1952ni}, capped by the Eddington rate, and involves both quasar- and radio-mode feedback regimes. Further details on the specific model of gas accretion and the AGN feedback are provided in~\citet{Hirschmann:2013qfl}. The Magneticum set includes a dmo counterpart for every hydro simulation. Additionally, a select number of these hydro simulations are paired with non-radiative simulations. The non-radiative simulations include gas hydrodynamics but no other subresolution physics.

In Table~\ref{tab:magnorig}, we report the relevant parameters regarding box size and mass resolution of the subset of Magneticum simulations we use as reference for our analysis. While the reference cosmology of the Magneticum simulations is WMAP7 ~\citep{WMAP:2010qai}, for the largest box here considered, Box 1a, the Magneticum suite also covers different cosmologies, as presented in Table~\ref{tab:magnmc}~\citep[see also][]{Singh:2019end}. Although cosmological parameters are varied, all these simulations keep the relevant parameters describing the subresolution models fixed. In addition, we also include four simulations that either vary the AGN feedback efficiency by $\pm 33$ percent, or assume a wind velocity of $500$ or $800$ ${\rm km\,s}^{-1}$, instead of the fiducial value of $350$ ${\rm km\,s}^{-1}$. 

\begin{table}
\begin{minipage}{\columnwidth}
\centering
\caption{Set of boxes from Magneticum simulations and halo-selection parameters used in this work. Column 1: name of the boxes; Column 2: size of the simulation box; Column 3: mass of the DM particle in each box; Column 4: minimum mass of the halos considered in our analysis; Column 5: the different physical models used to run the simulations.}
\resizebox{0.995\textwidth}{!}{%
{\renewcommand{\arraystretch}{1.25}
\setlength{\tabcolsep}{3pt}
\begin{tabular}{lcccccc}
 \hline\hline
Box & $L_\textrm{box}$ & $m_\textrm{DM}$ &$M_{\textrm{halo}, \textrm{min}}$ & \multicolumn{3}{c}{Physics} \\\cline{5-7}
& $(\textrm{Mpc})$ & ($M_\odot$) & ($M_\odot$) & hydro & dmo & non-radiative \\ \hline 
2 & $500$ & $9.8\times10^8$ & $4.3\times10^{13}$ & \checkmark & \checkmark & \tikzxmark \\
2b & $909$ & $9.8\times10^8$ & $4.3\times10^{13}$ & \checkmark\footnote{Box 2b ran only until $z=0.2$. } & \checkmark & \tikzxmark \\
1a & $1272$ & $1.9\times10^{10}$ & $4.3\times10^{14}$ &\checkmark&\checkmark & \checkmark\\
 \hline
\end{tabular}}}
\label{tab:magnorig}
\end{minipage}
\end{table}

\begin{table}
\begin{minipage}{\columnwidth}
\centering
\caption{Set of cosmologies covered by Magneticum Box 1a simulations used in this work.}
\resizebox{0.995\textwidth}{!}{%
{\renewcommand{\arraystretch}{1.25}
\setlength{\tabcolsep}{3pt}
\begin{tabular}{lccccccc}
 \hline\hline
Name & $\Omega_{\rm m}$ &       $\sigma_8$ & $h$ & $f_{\rm b}$ & \multicolumn{3}{c}{Physics} \\\cline{6-8}
     &            &            &       &       & hydro & dmo & non-radiative \\ \hline 
$C1 $ & $   0.153 $ & $ 0.614 $ & $     0.666 $ & $     0.267 $ & \checkmark & \checkmark & \checkmark\\
$C2 $ & $   0.189 $ & $ 0.697 $ & $     0.703 $ & $     0.241 $ & \checkmark & \checkmark & \tikzxmark\\
$C3 $ & $   0.204 $ & $ 0.739 $ & $     0.689 $ & $     0.214 $ & \checkmark & \checkmark & \tikzxmark\\
$C4 $ & $   0.200 $ & $ 0.850 $ & $     0.730 $ & $     0.208 $ & \checkmark & \checkmark & \tikzxmark\\
$C5 $ & $   0.222 $ & $ 0.793 $ & $     0.676 $ & $     0.190 $ & \checkmark & \checkmark & \tikzxmark\\
$C6 $ & $   0.232 $ & $ 0.687 $ & $     0.670 $ & $     0.178 $ & \checkmark & \checkmark & \tikzxmark\\
$C7 $ & $   0.268 $ & $ 0.721 $ & $     0.699 $ & $     0.168 $ & \checkmark & \checkmark & \tikzxmark\\
$C8 $\footnote{The reference WMAP7 cosmology.} & $   0.272 $ & $        0.809 $ & $   0.704 $ & $     0.168 $ & \checkmark & \checkmark & \checkmark\\
$C9 $ & $       0.304 $ & $     0.886 $ & $     0.740 $ & $     0.166 $ & \checkmark & \checkmark & \tikzxmark\\
$C10$ & $       0.301 $ & $     0.834 $ & $     0.708 $ & $     0.153 $ & \checkmark & \checkmark & \tikzxmark\\
$C11$ & $       0.342 $ & $     0.834 $ & $     0.708 $ & $     0.135 $ & \checkmark & \checkmark & \tikzxmark\\
$C12$ & $       0.363 $ & $     0.884 $ & $     0.729 $ & $     0.135 $ & \checkmark & \checkmark & \tikzxmark\\
$C13$ & $       0.400 $ & $     0.650 $ & $     0.675 $ & $     0.121 $ & \checkmark & \checkmark & \tikzxmark\\
$C14$ & $       0.406 $ & $     0.867 $ & $     0.712 $ & $     0.115 $ & \checkmark & \checkmark & \tikzxmark\\
$C15$ & $       0.428 $ & $     0.830 $ & $     0.732 $ & $     0.115 $ & \checkmark & \checkmark & \checkmark\\\hline
\end{tabular}}}
\label{tab:magnmc}
\end{minipage}
\end{table}

\subsubsection{Illustris TNG}

The Illustris TNG\footnote{\url{https://www.tng-project.org}} simulations were run using the AREPO code~\citep{Springel:2009aa}, which has a gravity solver similar to the P-Gadget3 one, with a hydro solver based on solving the Riemann problem on a moving mesh. This set of simulations represents an improvement ---in terms of the subresolution models adopted for star-formation and feedback models--- with respect to those adopted in the original Illustris simulations~\citep{Vogelsberger:2013eka}. It makes use of the same galaxy formation model with an improved kinetic AGN feedback model, a new parameterisation of galactic winds, and the addition of magnetic fields~\citep[see,][respectively]{2017MNRAS.465.3291W,Pillepich:2017jle,Pakmor:2011ht,Pakmor:2012xy,Pakmor:2013rqa}. 

For the purpose of the analysis presented in this paper, we use the largest box available; that is,\ the TNG300~\citep{2018MNRAS.477.1206N,Nelson:2017cxy,Springel:2017tpz,Pillepich:2017fcc,2019ComAC...6....2N} and its dmo counterpart. The simulated box size is $302.6$ Mpc and has a DM particle mass of $5.9\times10^7\,M_\odot$, the  assumed cosmological model being that of Planck 2015 ~\citep{Planck:2015fie}. We select all halos with mass higher than $1.0\times10^{13}\,M_\odot\,h^{-1}$. 

\subsubsection{Halo catalogues}

The halo catalogues used in this paper were obtained using the SUBFIND halo finder~\citep{Springel:2000qu,Springel:2020plp}, as presented in \cite{Dolag:2009}. This version of the code also includes gas and star particles. SUBFIND starts by establishing halo centres by running a parallel implementation of the 3D friends-of-friends~\citep[FOF; see e.g.][]{Davis:1985rj} algorithm and then allocating it to the position of the particle with the lowest potential. After that, it grows spheres around the centre and calculates the mass inside several radii, including a mean density $\Delta$ times the cosmic critical density. Throughout this paper, we only refer to the quantities with respect to the virial overdensity, $\Delta_{\rm vir}$, computed for the corresponding cosmology~\citep[see][]{Bryan:1997dn}.

\subsubsection{Matching algorithm}

For the purpose of the analysis presented in this paper, we need to identify the halo pairs corresponding to the same object identified in a hydro simulation and in its dmo counterpart. The matched catalogues are created by detecting the halo pairs from each of the hydro and dmo runs that are the closest based on the spatial proximity of their halo centres. Pairs that had a difference in mass of greater than $50$ percent were discarded. The same procedure was applied by~\citet{Castro:2020yes}, who stated that the matched catalogue completeness and purity are greater than $95$ percent for objects more massive than a few times $10^{13}\,M_\odot$. We independently validated the performance of the matching algorithm by comparing this position-based matching algorithm with a stricter matching algorithm based on matching objects that share more than $50$ percent of the DM particles.

\subsection{Baryonic correction on halo masses}\label{sec:corr}

The baryonic component can impact the spherical overdensity mass of galaxy clusters in three ways: due to adiabatic contraction, baryonic feedback, and the different dynamics between baryonic matter and DM. The adiabatic contraction is induced by radiative cooling of the diffuse gas. As the gas loses energy, its pressure support diminishes, leading to a fast collapse towards the bottom of the potential well, which in turn reacts by slightly contracting \citep[e.g.][]{Gnedin:2004cx}. This contraction is modulated by the rate of gas cooling. 

Concerning the feedback effects, these can cause sudden gas heating, and its subsequent displacement by expansion. Consequently, the DM halo responds by slightly expanding, with the extent of the expansion depending on the intensity of feedback. While feedback associated with SNe has been shown in simulations not to efficiently counteract adiabatic contraction on the scales of galaxy clusters and groups, AGN feedback is able to displace a large amount of gas. Consequently, the overall structure of the cluster is affected as the DM component is passively dragged by the baryonic component.

Lastly, another baryonic effect is the difference in accretion dynamics between DM and gas, with baryonic matter experiencing slower accretion due to the presence of shocks. This differential accretion can also affect the overall structure and mass of the galaxy clusters.

If baryonic effects happen slowly with respect to the cluster dynamical time and orbits are spherically symmetric, the angular momentum is an adiabatic invariant of the mass enclosed by the collapsing shell, which means there is conservation of the quantity $M(R)\,R$, where $M(R)$ is the total mass inside the radius $R$~\citep[see e.g.][]{1978AJ.....83.1050S,1980SvJNP..31..664Z,Blumenthal:1985qy,1987ApJ...318...15R}. Therefore, for a cluster-sized halo produced in the hydrodynamic simulation with virial mass $M_{\textrm{vir}, \textrm{hyd}}$ and virial radius $R_{\textrm{vir}, \textrm{hyd}}$ we can write
\begin{equation}
    M_{\Delta,{\rm dmo}} \, R_{\Delta,{\rm dmo}} = M_{\textrm{vir}, \textrm{hyd}} \, R_{\textrm{vir}, \textrm{hyd}}\,,
    \label{eq:adiabatic}
\end{equation}
where $M_{\Delta,{\rm dmo}}$ is the mass contained inside the sphere of radius $R_{\Delta,{\rm dmo}}$ centred on the dmo counterpart not affected by the baryonic physics. We note that, due to the baryonic impact on the halo profile, the overdensity threshold $\Delta$ in general will differ from the virial value. Defining $f_{\textrm{b}, \textrm{cosmic}}$ as the cosmic baryon fraction and $f_{\textrm{b}, \textrm{vir}}$ as the baryon fraction within the halo virial radius, then
\begin{equation}
    M_{\Delta,{\rm dmo}} = \frac{1-f_{\textrm{b}, \textrm{vir}}}{1-f_{\textrm{b}, \textrm{cosmic}}} \, M_{\textrm{vir}, \textrm{hyd}}\,.
    \label{eq:dmomass}
\end{equation}
Equation (2) is derived from the consideration that the DM contribution to the spherical overdensity mass in hydro-simulated halos, expressed as $(1-f_{\rm b, vir}), M_{\rm vir, hyd}$, equates to a mass that is $(1-f_{\rm b, cosmic})$ times larger than its dmo counterpart. This comparison considers that the dmo counterpart consists solely of a collisionless component, unlike the hydro-simulated halo, which includes both DM and baryonic matter. From Eqs.~\eqref{eq:adiabatic} and~\eqref{eq:dmomass}, it follows that
\begin{equation}
    \frac{R_{\Delta, {\rm dmo}}}{R_{\rm vir, hyd}} = \frac{1-f_{\rm b, cosmic}}{1-f_{\rm b, vir}} \, ,
    \label{eq:adiabatic2}
\end{equation}
and the overdensity $\Delta$ is defined as
\begin{equation}
    \Delta = \frac{3\, M_{\Delta,{\rm dmo}}}{4\,\pi\,R_{\Delta,{\rm dmo}}^3\,\rho_{\rm c}}\,,
    \label{eq:dmodelta}
\end{equation}
with $\rho_{\rm c}=3H^3/(8\pi G)$ the cosmic critical density and $H$ the Hubble parameter.

Several authors have studied the validity of the adiabatic approximation for the halo profile~\citep[see e.g.][]{Jesseit:2002tj,Gnedin:2004cx,Duffy:2010hf,Velmani:2022una}, showing that several of the assumptions made are violated; for instance, orbits can deviate strongly from sphericity due to major mergers. Furthermore, energy feedback effects can result in abrupt injection of gas kinetic energy in the cluster core, thus violating the quasi-static assumption. To take these effects into account, we modify Eqs.~\eqref{eq:dmomass} and~\eqref{eq:adiabatic2} as follows:
\begin{align}
    M_{\Delta,{\rm dmo}} &= \frac{1-f_{\rm b, vir} - \delta_f}{1-f_{\rm b, cosmic}} \, M_{\rm vir, hyd}\, ,\label{eq:quasiadiabatic1}\\
    \frac{R_{\Delta, {\rm dmo}}}{R_{\rm vir, hyd}} &= 1 + q\, \left( \frac{1-f_{\rm b, cosmic}}{1-f_{\rm b, vir} - \delta_{f}} - 1 \right) \, ,
    \label{eq:quasiadiabatic2}
\end{align}
where $q$ is a quasi-adiabatic parameter that controls the deviation from the adiabatic assumption~\citep[see][]{Schneider:2015wta,Paranjape:2021zia}, and $\delta_f$ accounts for the baryonic off-set, for which a halo containing $f_{\rm b, vir}=f_{\rm b, cosmic}-\delta_f$ has the same mass as its dmo counterpart. We note that zero correction corresponds to $\delta_f=0$ and $q=1$. It is important to emphasise that these two parameters, $q$ and $\delta_f$, both require calibration against simulations. This calibration is addressed in the following sections.

In summary, to derive the reconstructed dmo virial mass, we follow these steps:
\begin{enumerate}
    \item For a given mass $M_{\rm vir, hyd}$ we calculate the $M_{\rm \Delta, dmo}$ using Eq.~\eqref{eq:quasiadiabatic1}.
    \item We calculate $R_{\Delta, \rm dmo}$ using Eq.~\eqref{eq:quasiadiabatic2}.
    \item The dmo spherical overdensity threshold $\Delta$ is calculated from Eq.~\eqref{eq:dmodelta}.
    \item The reconstructed dmo virial mass $M_{\rm vir, rec}$ is estimated assuming that the dmo profile follows a Navarro-Frenk-White~\citep[NFW,][]{Navarro:1996gj} profile  with a concentration parameter given by~\citet{Diemer:2018vmz}.
\end{enumerate}
Therefore, the virial mass $M_{\rm vir, rec}(M_{\rm vir, hyd}, z, f_{\rm b, vir})$ can be reconstructed as a function of the virial mass of the hydro object $M_{\rm vir, hyd}$, its redshift $z$, and its baryonic fraction inside the virial radius $f_{\rm b, vir}$ following the steps outlined above. 

\subsection{Baryonic correction on the halo mass function}
\label{sec:hmfcor}

The HMF of hydro objects can be expressed as the convolution of the dmo counterpart with the probability distribution function $\mathcal{P}(M_{\rm vir, hyd}|M_{\rm vir, dmo})$:
\begin{equation}
    \frac{\diff n}{\diff M_{\rm vir, hyd}}  = \int \frac{\diff n}{\diff M_{\rm vir, dmo}} \, \mathcal{P}(M_{\rm vir, hyd} | M_{\rm vir, dmo}) \, \diff M_{\rm vir, dmo}\,,
    \label{eq:hmfconv}
\end{equation}
where $\mathcal{P}(M_{\rm vir, hyd}|M_{\rm vir, dmo})$ is the probability of an object with mass $M_{\rm vir, dmo}$ in the dmo simulation having a counterpart in hydro with mass $M_{\rm vir, hyd}$.

The convolution in Eq.~\eqref{eq:hmfconv} can be simplified  if the scatter between $M_{\rm vir, dmo}$ and $M_{\rm vir, hyd}$ is small enough that we can approximate the distribution with a Dirac delta function, as was done in~\citet{Castro:2020yes}. In this case, we can write
\begin{equation}
    \frac{\diff n}{\diff M_{\rm vir, hyd}}  = \frac{\diff n}{\diff M_{\rm vir, dmo}} \, \left|\frac{\diff M_{\rm vir, dmo}}{\diff M_{\rm vir, hyd}}\right|\,.
    \label{eq:hmfpred}
\end{equation}
If, in addition, we assume that the baryonic content in halos $f_{\rm b, vir}(M, z)$ of mass $M$ at redshift $z$ is fully predictive, we can make the following substitution in Eq.~\eqref{eq:hmfpred}:
\begin{equation}
    M_{\rm vir, dmo} \rightarrow M_{\rm vir, rec}(M_{\rm vir, hyd}, z, f_{\rm b, vir})\,,
\end{equation}
and use the model presented in Sect.~\ref{sec:corr} to compute the HMF in the hydro.

\subsection{Forecasting \Euclid's cluster count observations}
\label{sec:forecasts}

As was done by~\citet{Euclid:2022dbc} for the theoretical systematic effects concerning the calibration of the HMF, it is important to quantify how the uncertainties in the baryonic correction model impact the cosmological constraints for a realistic forecast of the cluster sample from the photometric Euclid Wide Survey~\citep{2022A&A...662A.112E}. 
We closely follow the forecasting procedure used in~\citet{Euclid:2021api} and~\citet{Euclid:2022dbc}, and refer to these works for further details. For brevity,  here we only highlight the main aspects: synthetic cluster survey data are generated considering a light cone covering $15\,000$ deg$^2$, with redshift range $z \in [0, 2]$~\citep{2011arXiv1110.3193L}. The number density of halos is sampled assuming the primary calibration for the HMF presented in~\citet{Euclid:2022dbc} with masses corrected by inverting the model described in Sect.~\ref{sec:corr} and assuming that the baryonic content inside the virial overdensity $f_{\rm b, vir} (M_{\rm vir, hyd}, z)$ follows a fiducial relation $f_{\rm b, fid} (M, z)$ extracted from either Magneticum or TNG300 simulations (see Appendix~\ref{app:content}).

Optical richness $\lambda$ is assigned to the halos of a given virial mass $M_{\rm vir}$ at redshift $z$ according to the relation
\begin{align}
        \langle \ln \lambda | M_{\rm vir}, z  \rangle =& \ln A_\lambda + B_\lambda \ln{\left(\frac{M_{\rm vir}}{3\times10^{14}\,h^{-1} M_\odot}\right)} \nonumber\\ 
        & + C_\lambda \ln{\left(\frac{E(z)}{E(z=0.6)}\right)}\,,
        \label{eq:lambda}
\end{align}
where $E(z)$ is the ratio of the Hubble parameter at redshift $z$ and at redshift $0$. We assume a richness range $\lambda \in [20,2000]$ and a log-normal scatter given by
\begin{equation}
        \sigma_{\ln \lambda | M_{\rm vir}, z}^2 = D_\lambda^2\, .
        \label{eq:lambda-var}
\end{equation}
We assume the same fiducial values as in~\citet{Euclid:2022dbc}, that is,\ $A_\lambda = 37.8$, $B_\lambda = 1.16$, $C_\lambda = 0.91$, and $D_\lambda = 0.15,$ based on the richness--mass relation presented by~\citet{DES:2015mqu}, and converting it to the virial mass definition assuming that halos follow a NFW profile and the mass--concentration relation given by~\citet{Diemer:2018vmz}. The adopted values are in agreement with the results presented by~\citet{Castignani:2016lvp}.

Finally, we assume a multivariate Gaussian likelihood with Poisson and sample variance fluctuations given by~\citet{Hu:2002we},~\citet{DES:2018crd}, and~\citet{Euclid:2021api}. To mimic the uncertainty on the baryon fraction relation and in our baryon correction model, we introduce two nuisance parameters, $\alpha$ and $\beta$. During the cosmological parameter inference, we assume that the baryon fraction is given by
\begin{equation}
    f_{\rm b, vir} (M_{\rm vir, hyd}, z) = (1+\alpha) \, f_{\rm b, fid} (M_{\rm vir, hyd}, z)\,,
\label{eq:alpha}
\end{equation}
and, similarly, we assume that the mass reconstruction given by our model relates to the fiducial dmo virial mass as
\begin{equation}
    M_{\rm vir, rec} (M_{\rm vir, hyd}, z, f_{\rm b, \, vir}) = 
    \frac{M_{\rm vir, dmo}}{1+\beta}\,.
\label{eq:beta}
\end{equation}
We assume Gaussian priors on $\alpha$ and $\beta$ and vary their standard deviation to assess the impact on the cosmological constraints.

\section{\label{sec:results}Results}

\subsection{\label{sec:cal}Model calibration}

A subset of simulations is leveraged in the calibration of the model, as described in Sect.~\ref{sec:methodology}. The choice of simulations is guided by a hierarchical approach aiming to ensure a robust calibration process that incorporates the physical effects incrementally.

For the calibration of the baryonic offset ($\delta_f$), we concentrate on the non-radiative versions of the Magneticum Box 1a. The rationale behind this choice is that these simulations serve as a fundamental calibration point for the baryon fraction, providing a baseline scenario wherein the other effects are assumed to balance to zero.

We used the
Magneticum hydrodynamic Box $2$ simulations to calibrate the quasi-adiabatic parameter ($q$). This introduces a more advanced level of simulation complexity, building upon the foundational understanding established through the calibration of $\delta_f$.

However, it is important to note that, in principle, both the baryonic offset and the quasi-adiabatic deviation could be influenced by the physics included in the simulations. Therefore, further calibration steps might be required when considering simulations with more detailed subresolution physical models.

\begin{figure*}
    \centering
    \includegraphics[width=\textwidth]{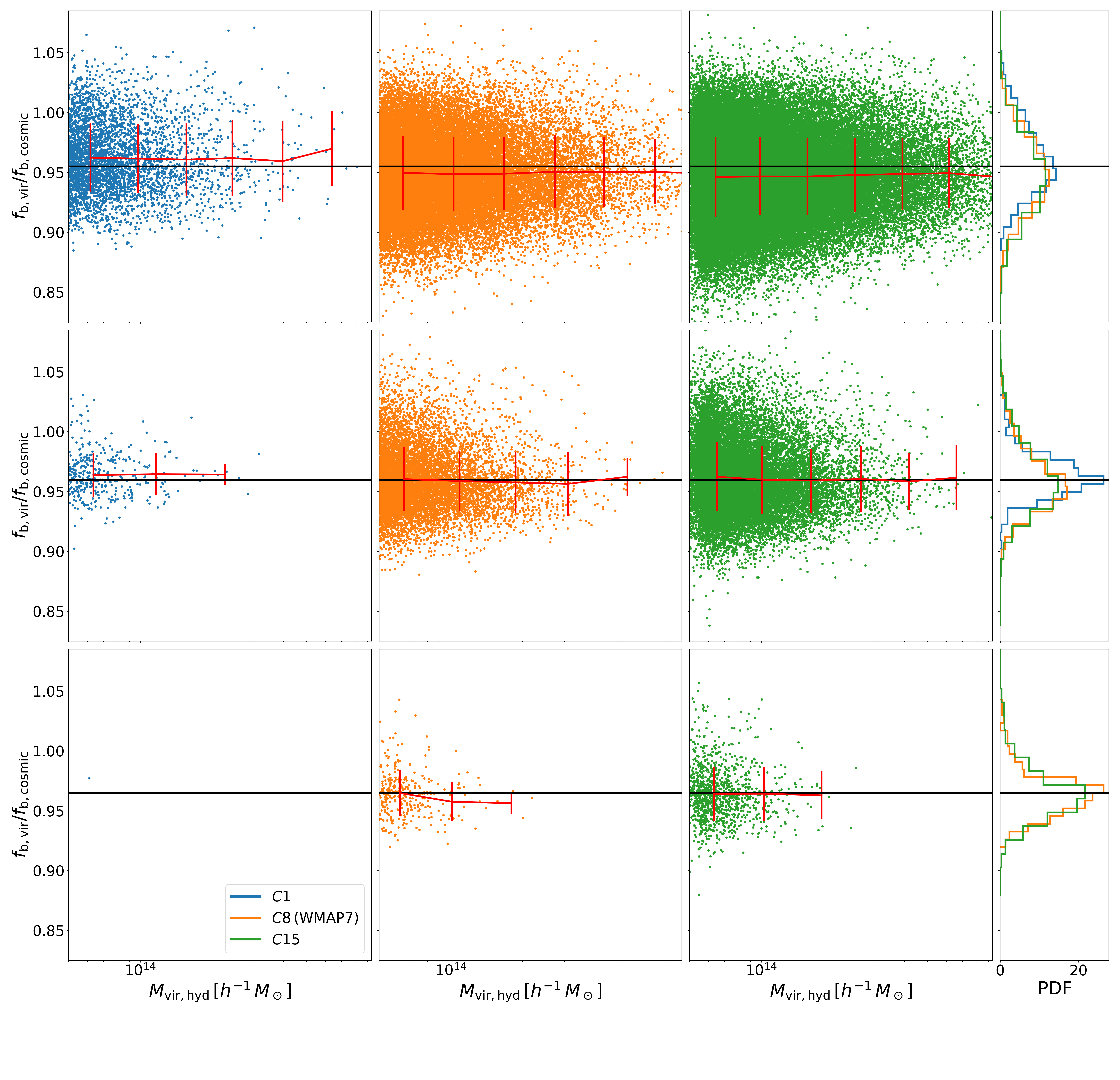}%\vspace{-1.25cm}
    \caption{Ratio of the baryonic fraction inside the virial radius with respect to the cosmic baryon fraction as a function of the virial mass for the three Magneticum adiabatic runs $C1$, $C8$, and $C15$ depicted in the three columns. Different rows correspond to the redshifts $z\in\{0.0, 1.0, 2.0\}$. The mean relation, as given by Eq.~\eqref{eq:mean-baryons} as a function of redshift, is depicted as the horizontal black line while the red lines correspond to the measured mean and unbiased standard deviations in mass bins. The rightmost column shows the distribution of these ratios for the three simulations.}
    \label{fig:adiabatic-fb}
\end{figure*}
\subsubsection{Baryonic content in non-radiative simulations}

In Fig.~\ref{fig:adiabatic-fb}, we present the ratio between the baryonic fraction inside the virial radius and the cosmic baryon fraction, as a function of the virial mass for the three Magneticum simulations for which the non-radiative version is available: $C1$, $C8$, and $C15$ reported in the three columns. The result for the redshifts $z\in\{0.0, 1.0, 2.0\}$ is presented in the different rows. The red lines correspond to the mean and unbiased standard deviations measured for the  mass bins. We note that the mean baryonic content inside the virial radius scatters around the following relation despite the background cosmological parameters (marked by the horizontal black line):
\begin{equation}
    \left\langle \frac{f_{\rm b, vir}}{f_{\rm b, cosmic}} \right\rangle = 1 - \left( 0.045 - 0.005\,z \right)\,.
    \label{eq:mean-baryons}
\end{equation}
The above relation was obtained by fitting a linear relation for the mean relation as a function of redshift.

\begin{figure*}
    \centering
    \includegraphics[width=\textwidth]{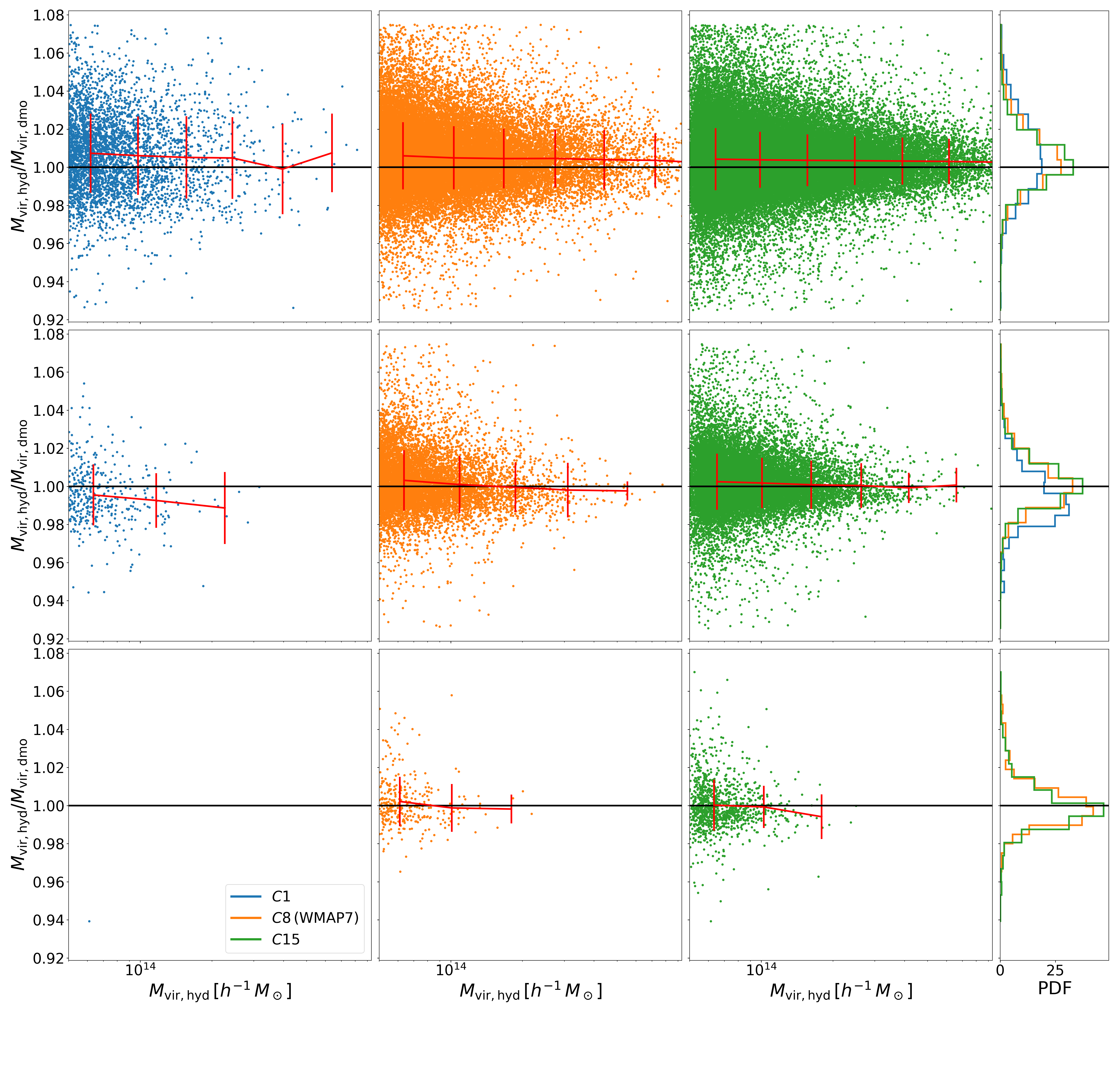}%\vspace{-1.25cm}
    \caption{Ratio of the virial mass in the non-radiative run with respect to the matched dmo counterpart for the three Magneticum non-radiative runs. Different rows correspond to the redshifts $z\in\{0.0, 1.0, 2.0\}$. The unity relation is depicted as the horizontal black line, while the red lines correspond to the measured mean and unbiased standard deviations in mass bins.}
    \label{fig:adiabatic-mbias}
\end{figure*}

Conversely, in Fig.~\ref{fig:adiabatic-mbias}, we present the ratio of the virial mass in the non-radiative run with respect to the matched dmo counterpart. Despite the missing baryonic content shown in Fig.~\ref{fig:adiabatic-fb}, we observe that halos have the same average mass in the dmo and non-radiative simulations. This leads to the following relation for $\delta_f$ in Eqs.~\eqref{eq:quasiadiabatic1} and~\eqref{eq:quasiadiabatic2}:
\begin{equation}
    \delta_f = (0.045 - 0.005\,z)\,f_{\rm b, cosmic}\,.
    \label{eq:deltaf}
\end{equation}

\subsubsection{Quasi-adiabatic response}

In order to determine the value of the quasi-adiabatic parameter $q$ that  appears in Eq.~\eqref{eq:quasiadiabatic2} for the model presented in Sect.~\ref{sec:methodology}, we proceed in the following way. We bin the halo catalogues of Box $2$ of the Magneticum set at five redshifts, $z\in\{0.00, 0.25, 0.50, 1.00, 2.00\}$ in $\log_{10} M_{\rm vir, hyd}/(h^{-1}\,M_\odot)$ with a bin width of $0.3$ dex.
Within each mass bin, we compute the mean and the unbiased standard deviation of the ratio of the virial mass in the hydrodynamic and in the dmo simulations, along with the mean of the ratio of the inferred mass and the true dmo mass. Then, the $\chi^2$ between the quasi-adiabatic model and the results from simulations is computed as
\begin{equation}
    \chi^2 = \sum_{i, j} \left(\frac{\langle M_{\rm vir, rec}/M_{\rm vir, dmo} \rangle - \langle M_{\rm vir, hyd}/M_{\rm vir, dmo} \rangle}{\sigma_{i,j}} \right)_{i, j}^2\,,
\end{equation}
where $i$ and $j$ run over all mass bins and redshifts. In the above equation, $\sigma_{i,j}$ is the error in the binned statistics given by
\begin{equation}
\sigma_{i,j} = \sqrt{\frac{{{\rm{var}}\left(\frac{M_{\rm vir, hyd}}{M_{\rm vir, dmo}}\right)}_{i,j}}{N_{i,j}} + 0.003^2}\,,
\end{equation}
where ${\rm var}$ is the unbiased variance estimation of the ratio of the virial mass in the hydrodynamic and in the dmo simulations divided by the number of objects in the bin. To this, we add a constant fixed at 0.003  in quadrature. This constant is introduced to adjust for the simplicity of our error modelling, which is estimated directly from the data, possibly understating the true uncertainties in the relation. The addition of this constant ensures a resulting $\chi^2$ of the order of the unity.

\begin{figure}
    \centering
    \includegraphics[width=\columnwidth]{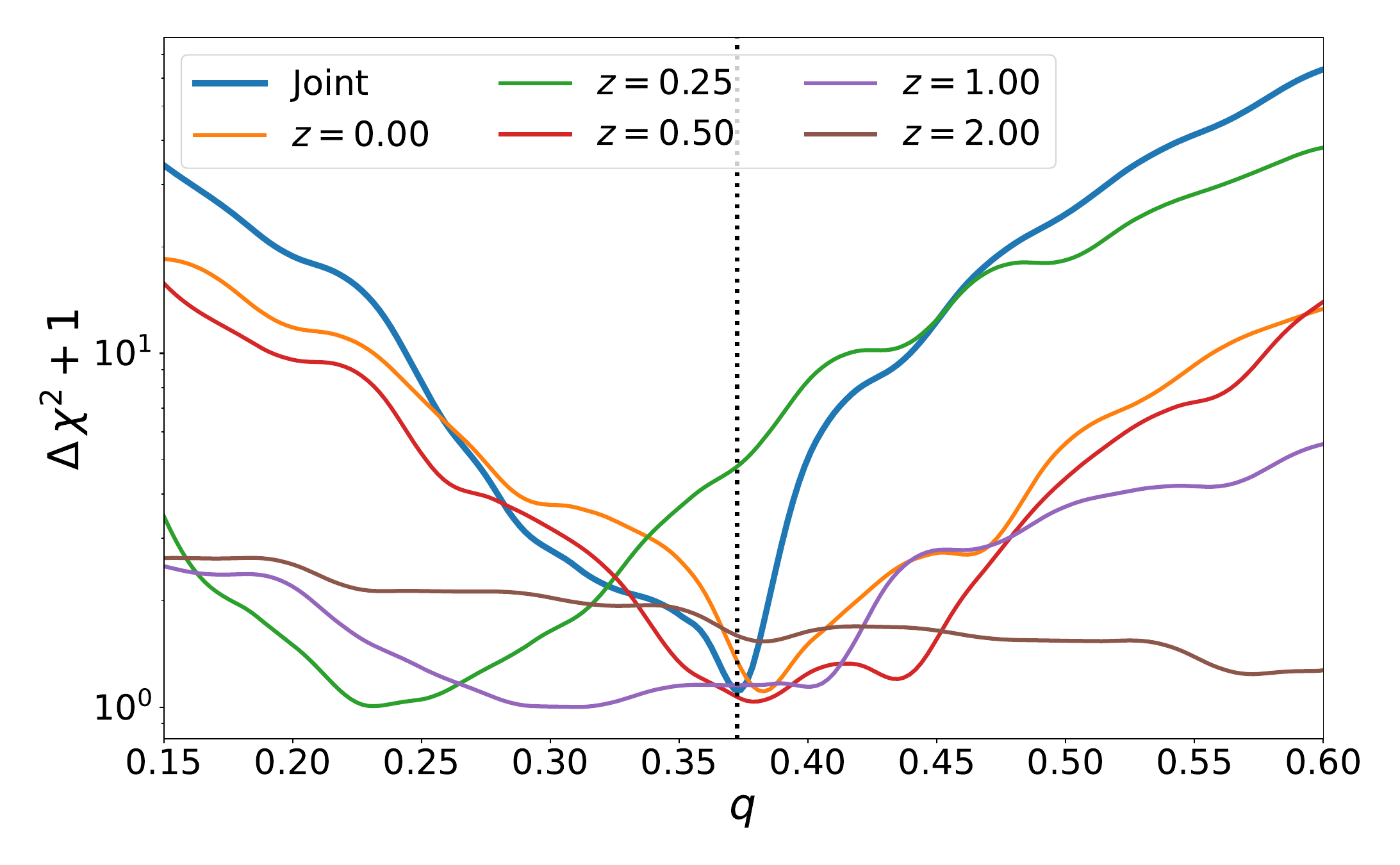}
    \caption{Variation of the $\chi^2$ with respect to its minimum, as a function of the quasi-adiabatic parameter $q$ for the model presented in Sect.~\ref{sec:methodology}. We present the results for the joint analysis of five redshifts $z\in\{0.00, 0.25, 0.50, 1.00, 2.00\}$ from the Magneticum Box $2$ simulation, as well as the results at each redshift. The best fit for $q=0.373$ is marked with the vertical dashed line.}
    \label{fig:fit-q}
\end{figure}

In Fig.~\ref{fig:fit-q}, we plot the values of the $\Delta \chi^2$ as a function of $q$, both for single redshifts and for the analysis based on combining all redshifts. The $\Delta \chi^2$ is defined as:
\begin{equation}
\Delta \chi^2 = \chi^2 - \chi^2_{\rm best-fit}\,,
\end{equation}
where $\chi^2_{\rm best-fit}$ corresponds to the best fit of the joint analysis. We observe that all redshifts prefer significant deviations from the adiabatic prediction ($q=1$). All redshifts present a minimum around the joint analysis minimum $q=0.373$ (presented as the vertical black dotted line). Exceptions occur at $z=0.25$ and $z=2.0$ that have their best fit shifted to $q\simeq 0.23$ and $q\simeq 0.60$, respectively. Still, assuming Gaussianity, those shifts correspond to only a $2\sigma$ deviation for $z=0.25$ and less than $1\sigma$ for $z=2.0$. Given the lack of strong statistical significance, we interpret this deviation as likely a result of statistical variation rather than a reflection of any substantive physical process at this particular redshift. Therefore, in the following, we assume the joint analysis best fit for $q$ to hold at all redshifts.

\subsection{\label{sec:validation} Validation of the model}

\begin{figure*}
    \centering
    \includegraphics[width=\textwidth]{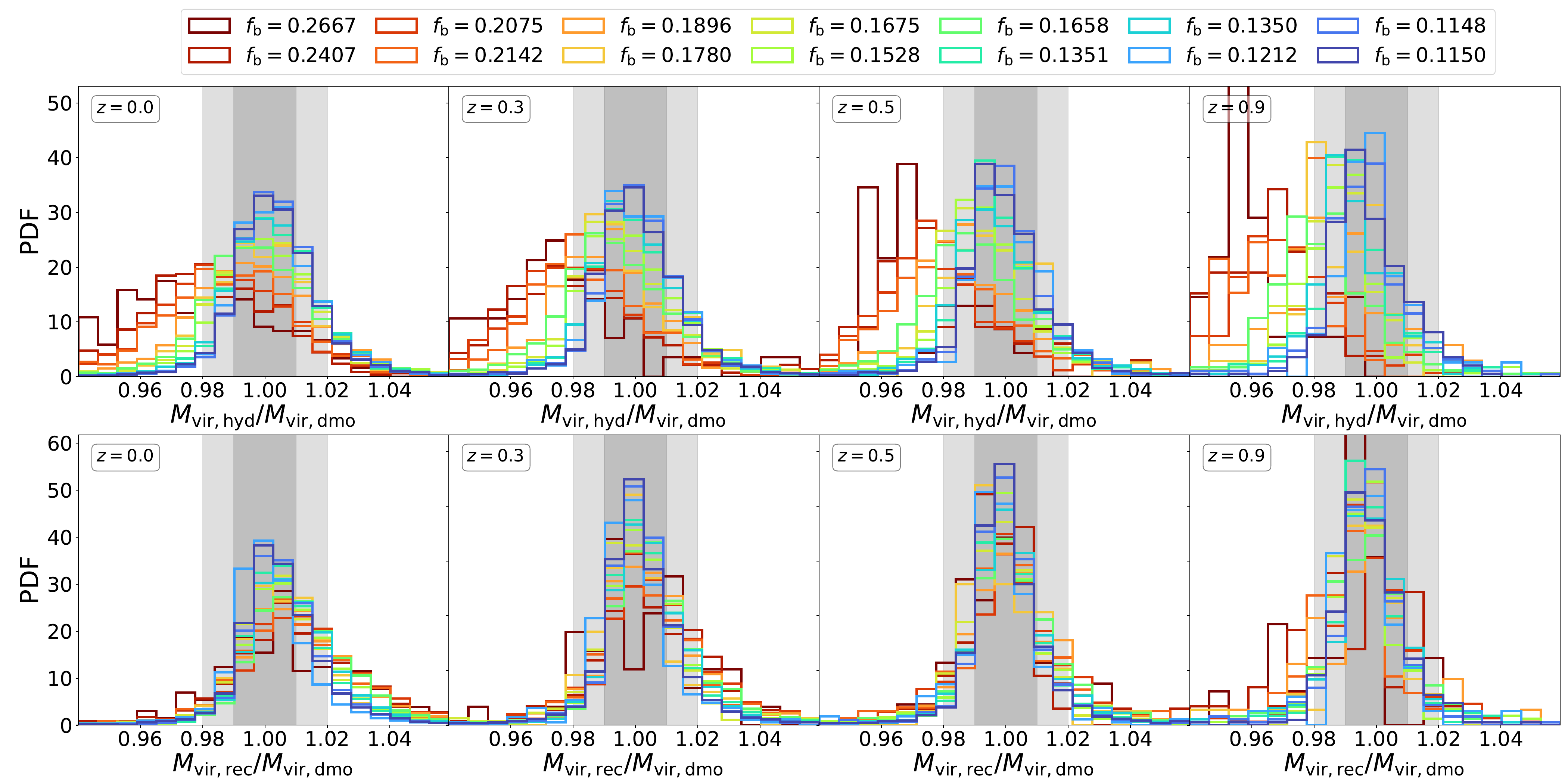}
    \caption{Ratio between either the hydrodynamic mass (\emph{top} panels) or the reconstructed virial mass (\emph{bottom} panels) and the dmo counterpart on the subset of Magneticum Boxes 1a for the simulations with different cosmologies (see Table~\ref{tab:magnmc}). Different columns correspond to the four redshifts $z\in\{0.0, 0.3, 0.5, 0.9\}$ and different colours correspond to different cosmic baryon fractions $f_{\rm b}$. The filled dark-grey and light-grey areas mark the $1$ and $2$ percent halo mass variation induced by baryonic effects, respectively.}
    \label{fig:mbias-pdf-1}
\end{figure*}

\begin{figure}
    \centering
    \includegraphics[width=\columnwidth]{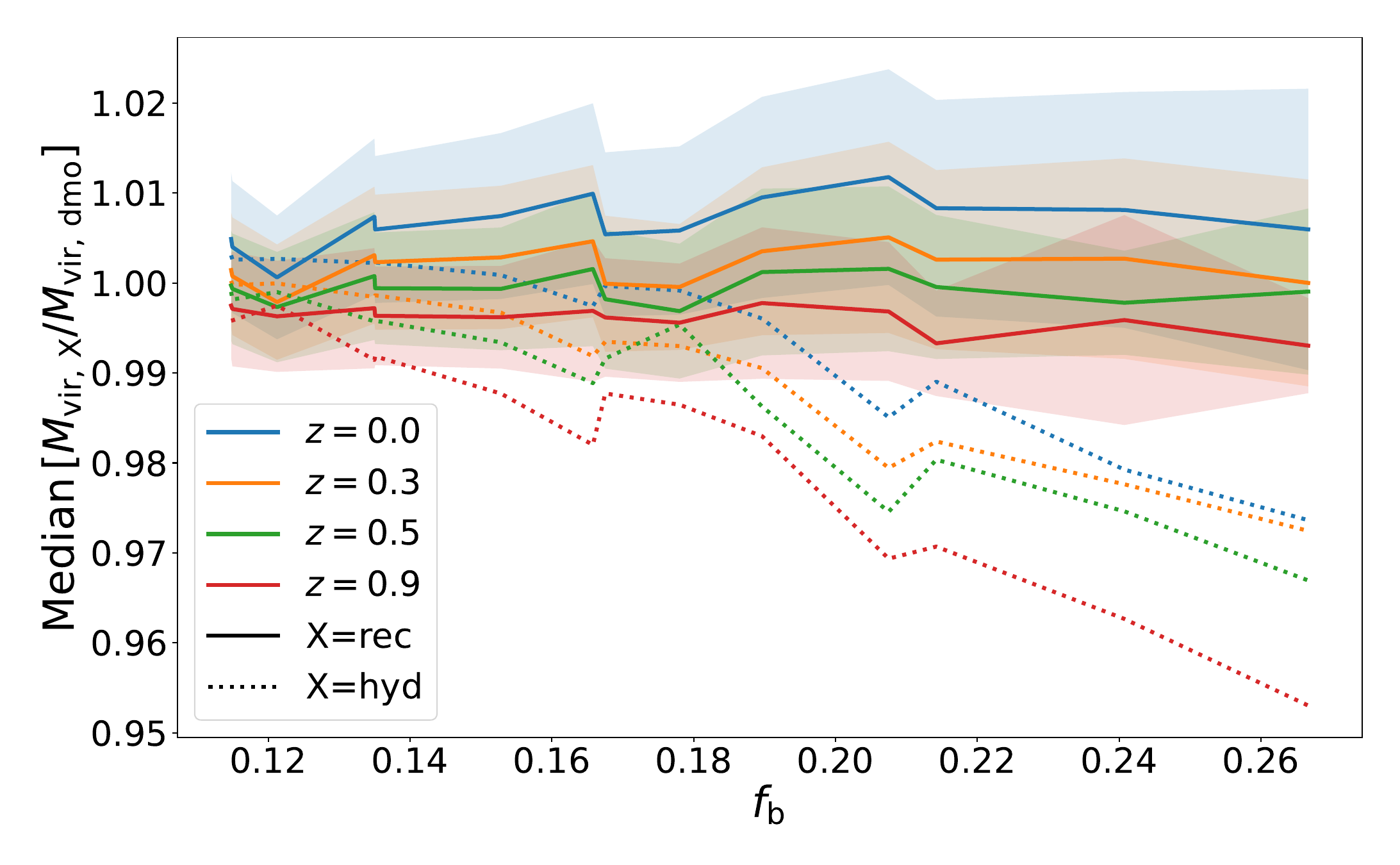}
    \caption{Median of the distributions shown in Fig.~\ref{fig:mbias-pdf-1} as a function of the cosmic baryon fraction $f_{\rm b}$ for the ratio between either the hydrodynamic mass (dotted lines) or the reconstructed virial mass (full lines) and the dmo counterpart on the subset of Magneticum Boxes 1a for the simulations with different cosmologies (see Table~\ref{tab:magnmc}). Different colours correspond to the four redshifts $z\in\{0.0, 0.3, 0.5, 0.9\}$. The filled regions corresponds to one median absolute deviation around the median.}
    \label{fig:mbias-median}
\end{figure}

\begin{figure*}
    \centering
    \includegraphics[width=\textwidth]{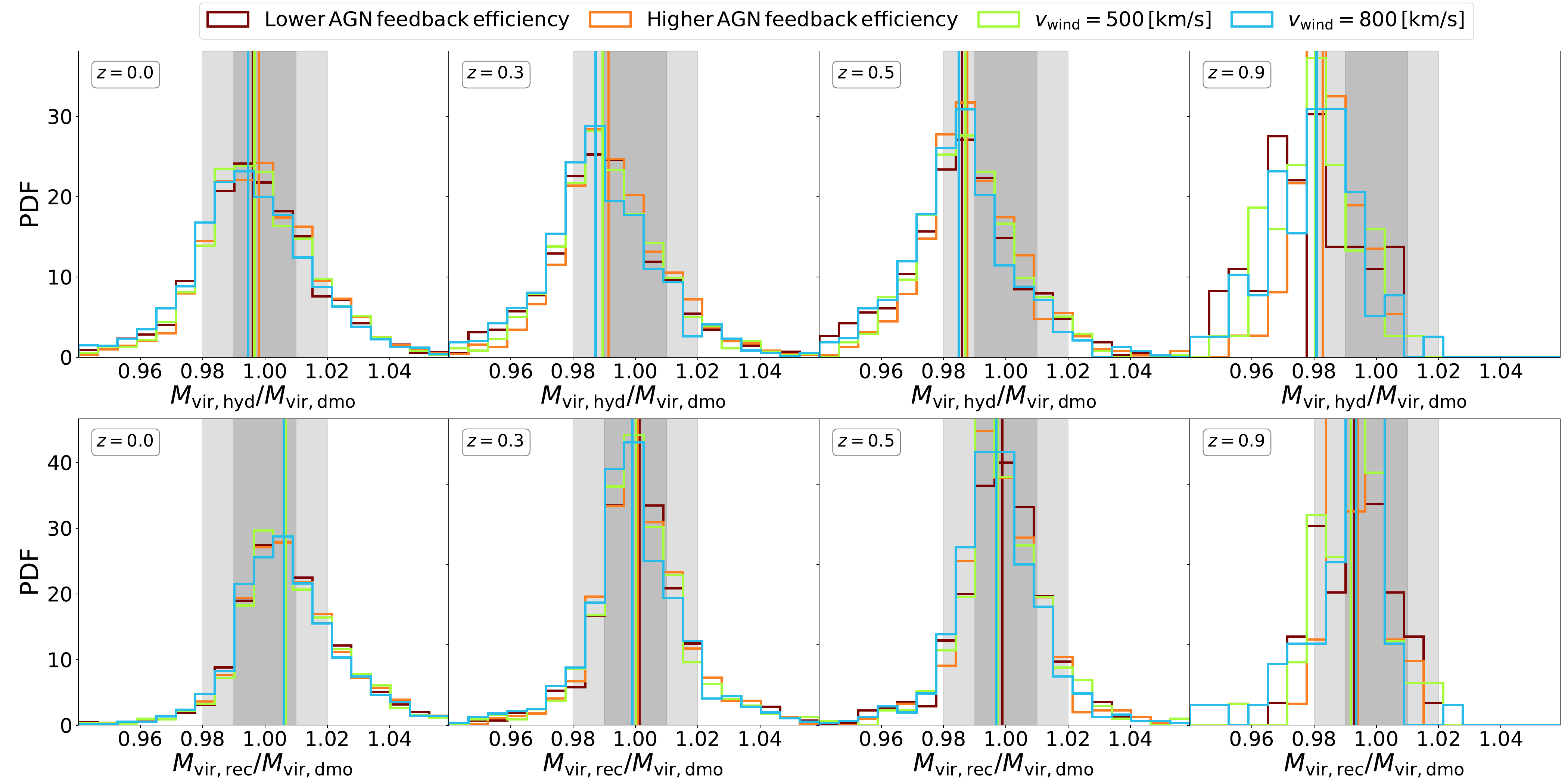}
    \caption{Similar to Fig.~\ref{fig:mbias-pdf-1} but for the simulations with $C8$ (WMAP7) background cosmology that assume different values of the parameters describing SN and AGN feedback.}
    \label{fig:mbias-pdf-2}
\end{figure*}

In Fig.~\ref{fig:mbias-pdf-1}, we present the ratios of the original halo masses in hydro simulations (\emph{top} panel) and the reconstructed halo masses (\emph{bottom} panel), both to the dmo counterpart. Results are shown for the subset of Magneticum Boxes 1a that assume different cosmological parameters (see Table~\ref{tab:magnmc}), at four redshifts, $z\in\{0.0, 0.3, 0.5, 0.9\}$. The medians of the corresponding distributions are shown in Fig.~\ref{fig:mbias-median}. The bottom panel presents the model's performance in recovering the dmo mass for the simulations with different cosmologies colour-coded by the baryon fraction. 

As expected, the median value of the ratio between hydro and dmo halo masses shows significant deviations from unity, in a way that becomes more significant as the value of the baryon fraction increases. On the other hand, we see from the bottom panels that our model, on
average, allows us to correctly recover the halo masses from the dmo simulations. Quite remarkably, this accuracy is independent of the cosmological model adopted in the simulations, as the correlation of the net effect with the baryonic fraction is absent in the reconstructed mass. In general, dmo masses are recovered by our model with subpercent accuracy. Possible exceptions are represented by the simulations with the largest cosmic baryon fraction at $z=0$, for which our model seems to slightly over-predict the reconstructed mass by roughly $1$ percent. Lastly, we observe that subdominant to the model target accuracy of $1$ percent, there is a correlation between the model performance and the redshift, with the median of the ratio decreasing with increasing redshift. This suggests that there is a feature missing in our model, such as a redshift-dependent quasi-adiabatic factor reflecting that at higher redshift, the effect is driven by radiative cooling, while at low redshift AGN feedback is the main agent~\citep[see][]{Castro:2020yes}. An increase in the sophistication of the model to push our model accuracy to the subpercent level is left for further investigation in future work.

Similarly to Fig.~\ref{fig:mbias-pdf-1}, in Fig.~\ref{fig:mbias-pdf-2} we present the performance of the model for the simulations with $C8$ (WMAP7) background cosmology, but assuming different values for the parameters that define the efficiency of feedback from both AGN and SN. Vertical lines denote the median of the corresponding distribution. In particular, we varied {\em (a)} the efficiency of the AGN feedback by changing the value of the \texttt{BlackholeFeedbackFactor} parameter from 0.1 to 0.2 (default value was 0.15), and {\em (b)} the value of the velocity of the SN-driven galactic outflows in the model by \cite{Springel:2003} from 350 ${\rm km\,s}^{-1}$ to 500 ${\rm km\,s}^{-1}$ and 800 ${\rm km\,s}^{-1}$, respectively. In general, we note that the effect of changing the parameters regulating stellar and AGN feedback is relatively small, especially at low redshift. In any case, the accuracy with which our model recovers the halo masses in the dmo simulation is not degraded by changing the feedback parameters. The difference between the vertical lines is always smaller for the reconstructed mass than when considering the actual masses from the hydro simulations.

\begin{figure*}
    \centering
    \includegraphics[width=\textwidth]{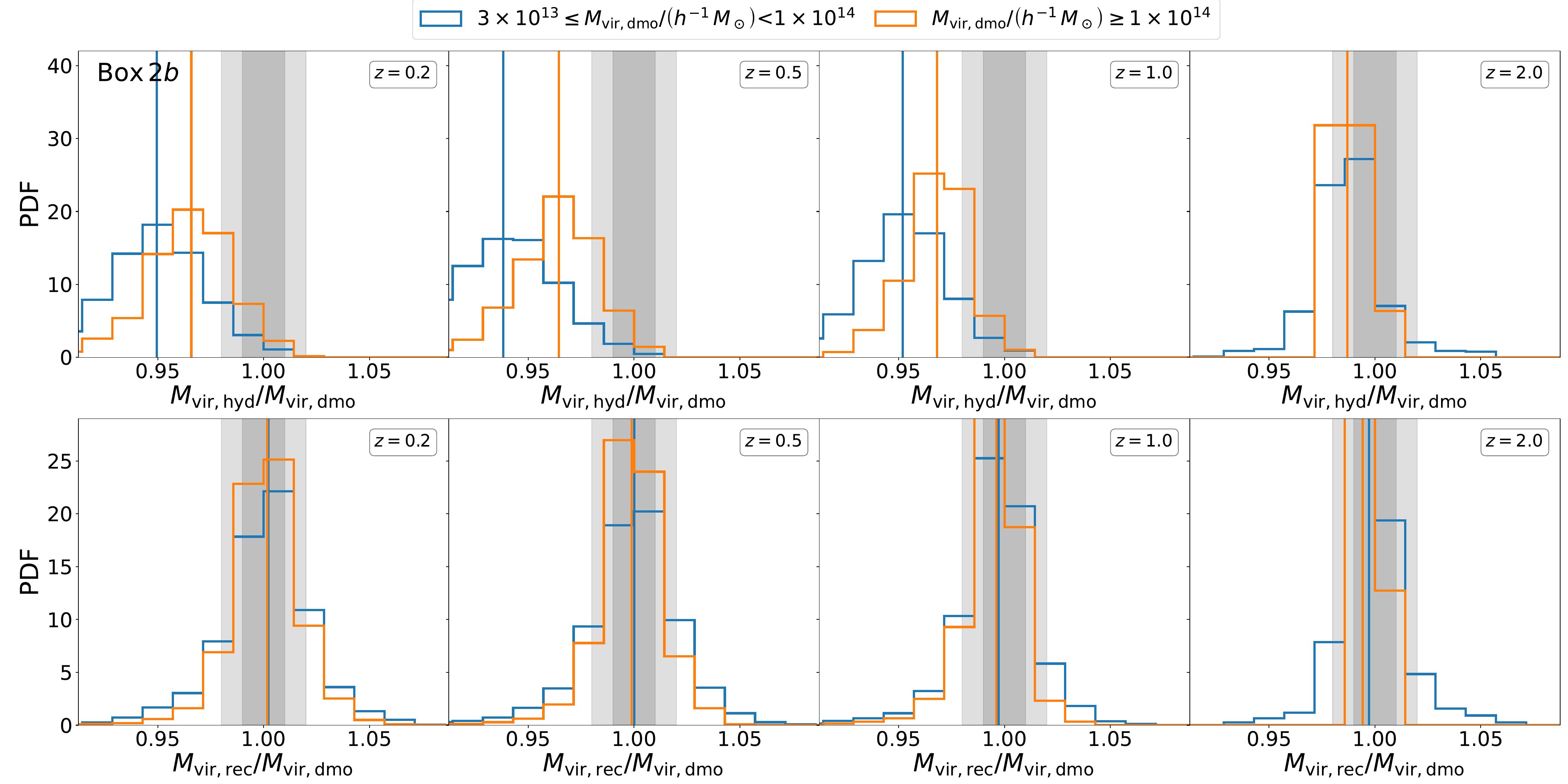}
    \caption{Ratio of the hydrodynamic mass (\emph{top} panels) and of the reconstructed virial mass (\emph{bottom} panels) to the dmo counterpart on the Magneticum Boxes 2b. The different columns present the results for $z\in\{0.2, 0.5, 1.0, 2.0\}$. We show the results for two mass regimes: $3\times 10^{13} \leq M_{\rm vir, dmo}/(h^{-1} M_\odot) < 10^{14} $ (blue histograms) and $M_{\rm vir, dmo}/(h^{-1} M_\odot) \geq 10^{14} $ (orange histograms).}
    \label{fig:mbias-box2b-pdf}
\end{figure*}

\begin{figure*}
    \centering
    \includegraphics[width=\textwidth]{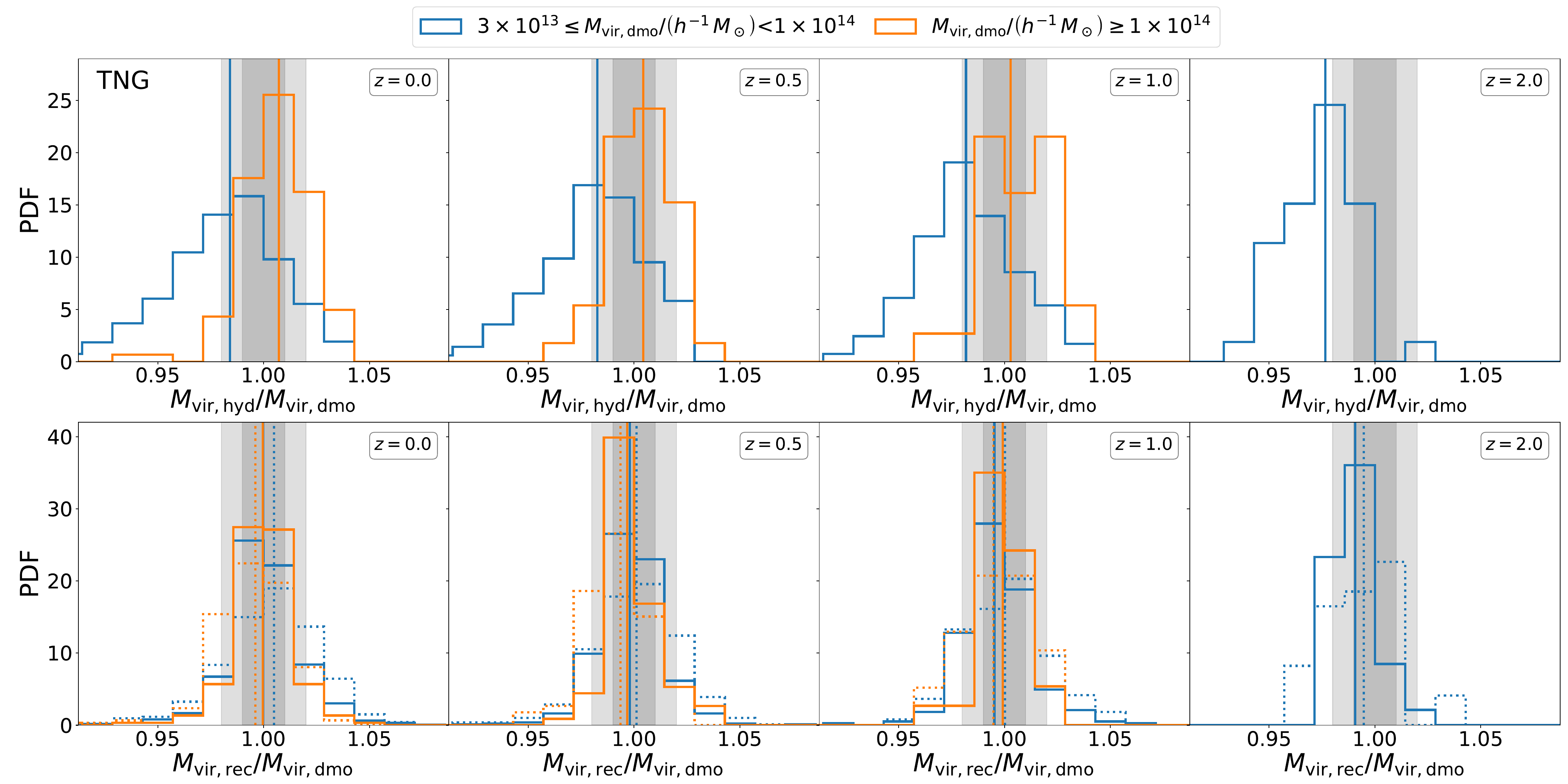}
    \caption{Similar to Fig.~\ref{fig:mbias-box2b-pdf} but for the TNG300 simulation.}
    \label{fig:mbias-tng-pdf}
\end{figure*}

Similarly to Figs.~\ref{fig:mbias-pdf-1} and~\ref{fig:mbias-pdf-2}, in Figs.~\ref{fig:mbias-box2b-pdf} and~\ref{fig:mbias-tng-pdf} we present the performance of the model on the Magneticum Box $2$b and on the TNG300 simulations, respectively. The different columns present the results for $z\in\{0.2, 0.5, 1.0, 2.0\}$ for Magneticum and $z\in\{0.0, 0.5, 1.0, 2.0\}$ for TNG300. 
We note that considering Box 2b instead of Box 1a of the Magneticum set allows us to validate the robustness of our method when increasing resolution. At the same time, the analysis of the TNG300 box allows us to further stress the validity of our model at an even higher resolution and for a substantially different hydrodynamical solver, star-formation model, and implementation of both stellar and AGN feedback. We present the results for two mass regimes: $3\times 10^{13} \leq M_{\rm vir, dmo}/(h^{-1} M_\odot) < 10^{14} $ and $M_{\rm vir, dmo}/(h^{-1} M_\odot) \geq 10^{14} $. We refer to the former as the group regime and the latter as the regime of galaxy clusters. We note that, due to the smaller size of the TNG300 box, this simulation does not contain cluster-sized halos at $z=2$. Both simulations predict a stronger impact on the group regime than on the cluster regime; this is expected as the potential well is shallower  in the
former regime than in the latter, and therefore the baryonic depletion requires less feedback. In general, TNG300 predicts a smaller impact of baryonic effects on halo masses in the cluster regime than the Magneticum Box 2b simulation. In any case, our model performs equally well in both simulations and mass regimes. The smaller impact predicted by the TNG300 simulations is due to the fact that Magneticum and TNG300 simulations predict a similar baryonic content for virial halos at $z=1$ as can be seen in Fig.~\ref{fig:fb-sims}. However, the cosmic baryon fraction assumed in the Magneticum reference simulation is $\sim 7$ percent higher, and therefore it can be inferred that the Magneticum simulations exhibit a more pronounced gas depletion and more active feedback mechanisms compared to those in TNG. This results in a greater impact on halo mass in Magneticum due to a more significant displacement of gas.

\subsection{\label{sec:hmf}Robustness of the model to the assumed baryon fraction relation}

Above, we demonstrate the performance of our model in reconstructing $M_{\rm vir, dmo}$ under the assumption that the baryon fraction for each halo is known. Additionally, it is of interest to evaluate the performance of the model when employing an averaged baryon fraction–mass relation across a sample, as an alternative approach. This approach is arguably more realistic for implementation in cosmological analyses.

\begin{figure}
    \centering
    \includegraphics[width=0.975\columnwidth]{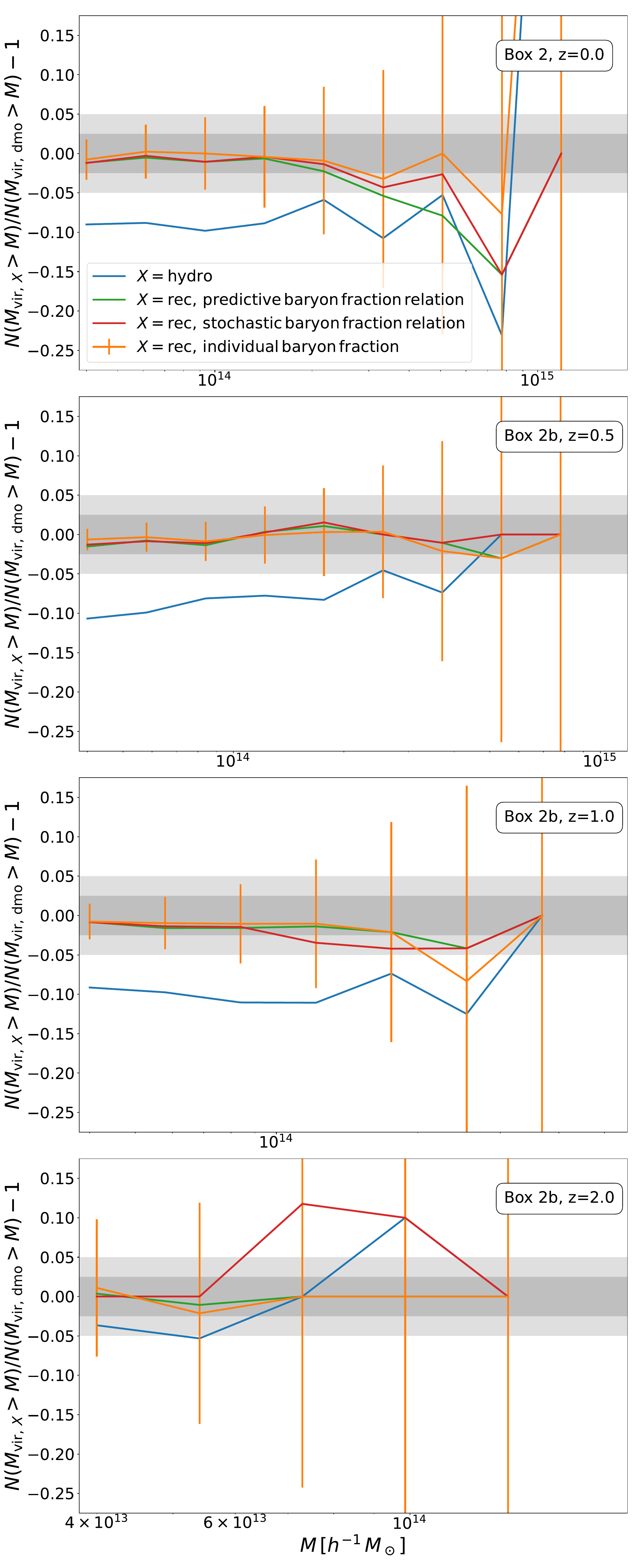}
    \caption{Ratio of the cumulative halo abundance in the hydro simulation to the halo abundance of virial halos in the dmo simulation. The results are shown for the simulated virial mass and the prediction of our model based on the individual baryon fractions, a predictive baryon fraction relation, and a stochastic baryon fraction relation. Error bars are displayed for the first case, assuming uncorrelated Poisson errors. Different panels correspond to redshifts $z\in\{0.0, 0.5, 1.0, 2.0\}$ using Magneticum Box 2b, except for $z=0.0$, where Box 2 is used. The grey regions correspond to $2.5$ and $5$ percent.}
    \label{fig:hmf-rec}
\end{figure}

In Fig.~\ref{fig:hmf-rec}, we present the ratio between the cumulative halo abundance assuming different masses in the hydro simulation and the halo abundance of virial halos in the dmo simulation. We present the results for the virial mass of the hydro simulation as well as the virial prediction of our model when taking into account the following assumptions: 
\begin{enumerate}
    \item That the baryon fraction of individual objects is known.
    \item A predictive baryon fraction relation given by Eq.~\eqref{eq:fb-sims} with parameters shown in Table~\ref{tab:par} for the Magneticum set.
    \item A stochastic baryon fraction with mean given by Eq.~\eqref{eq:fb-sims} with parameters shown in Table~\ref{tab:par} for the Magneticum set and with a scatter of 0.17 dex.
\end{enumerate}
The third case addresses the possibility that the scatter of the cluster gas fraction in the simulation is underestimated compared to observational data. We assume the scatter of $0.17$ dex reported by~\citet{Andreon:2017rih} for the gas fraction. To increase the readability of the  plot, we only present error bars for the first case, assuming uncorrelated Poisson errors for the abundances. Different panels correspond to the redshifts $z\in\{0.0, 0.5, 1.0, 2.0\},$ and for all panels we use Magneticum Box 2b, except for $z = 0.0,$ where we use Box 2, as Box 2b was not run to this redshift.

In Fig.~\ref{fig:hmf-rec}, the baryonic physics tends to produce
halos that are less massive than their dmo counterparts, as previously discussed by~\cite{Castro:2020yes}. Therefore, at fixed mass value, the abundance of halos in a hydro simulation is lower than that derived from a dmo simulation (blue lines). Conversely, converting $M_{\rm vir, hyd}$ to the corresponding dmo virial masses, the abundances of the two simulations match (green and orange lines). Regarding the performance of our model, we note that using a collective relation (red and green lines) instead of the individual object baryon fraction (orange line) does not significantly degrade the mass reconstruction. This is neither obvious nor expected a priori, as non-linear functions do not commute with the median operation. The robustness of the performance comes from the tightness of the baryon fraction relation at fixed mass (see Fig.~\ref{fig:fb-sims}), and ensures the applicability of our model in cluster cosmology studies, even if we assume a larger variance for the baryon fraction distribution. Therefore, the approximations that lead to the simplification of Eq.~\eqref{eq:hmfconv} to Eq.~\eqref{eq:hmfpred} are validated and do not statistically affect the model's performance.

\section{\label{sec:euclid}Impact of baryonic effects on a \Euclid-like cluster abundance analysis}

In this section, we quantify the biases in deriving the posteriors on the cosmological parameters $\Omega_{\rm m}$ and $\sigma_8$ caused by neglecting the baryonic effects on halo masses. Specifically, we assess these biases using a fully predictive baryon fraction relation measured from either the Magneticum or the TNG300 simulations when creating the synthetic catalogue (see Appendix~\ref{app:content} for more details). Following this, we evaluate the efficacy of our method in recovering halo masses in dmo simulations.

 We perform the forecast following the methodology described in Sect.~\ref{sec:forecasts}. For the likelihood analysis, we assume flat priors on the cosmological parameters and Gaussian priors on the mass--observable parameters of Eqs.~\eqref{eq:lambda} and~\eqref{eq:lambda-var} with the mean given by the fiducial values and with a rms\ of 1 and 3 percent, respectively. The likelihood sampling is performed following the ensemble slice sampling method~\citep[see][]{karamanis2020ensemble} implemented in ZEUS~\citep[see][]{karamanis2021zeus}. 

Firstly, we assess the tension between the inferred cosmological parameters and the fiducial ones after marginalising over the other parameters, as quantified by the index of inconsistency \citep[IOI;][]{Lin:2017ikq}, which is calculated as
\begin{equation}
    {\rm IOI} = \frac{\boldsymbol{\delta}^{\rm t}\,\boldsymbol{\Sigma}^{-1}\,\boldsymbol{\delta}}{2}\,.
\end{equation}
In the above expression, $\boldsymbol{\delta}$ is the two-dimensional difference vector between the best-fit values and the fiducial values of the $\Omega_{\rm m}$ and $\sigma_8$ cosmological parameters, while $\boldsymbol{\Sigma}$ is the covariance matrix between these parameters, which we assume to be Gaussian distributed. We calculate the IOI when ignoring the baryonic impact, assuming the cosmological parameters and the virial baryon content relation from the TNG and the Magneticum simulations. 

In Table~\ref{tab:IOI&FOM}, we present the summary statistics for the forecast of the impact of the treatment of the baryonic effects on halo masses. We present the results for the constraints on $\Omega_{\rm m}$ and $\sigma_8$ to be obtained from the cluster counts analysis of the forthcoming \Euclid survey. Ideally, the IOI should be kept below 1 to ensure that any correlations with other systematic effects do not amplify tensions in the final results~\citep{Euclid:2023wdq}.

For both Magneticum and TNG simulations, we observe an increasing impact of the baryonic effects on halo masses  as stronger priors on the mass--observable scaling relation are assumed. This is an expected result, as a tighter prior on the scaling relation results in tighter constraints on the cosmological parameters, making cosmological posteriors more sensitive to systematic effects. While ignoring baryonic effects in the Magneticum simulations would lead to an IOI of $4.1$ and $2.0$ for $1$ and $3$ percent priors, respectively, for TNG300 these values reduce to $1.8$ and $0.9$. 

The reasons for the smaller impact of baryons predicted by TNG with respect to Magneticum are two fold.
\begin{enumerate}
    \item Baryon content in TNG300: As mentioned in Section~\ref{sec:validation}, the baryon content in massive clusters observed in TNG300 more closely aligns with the cosmic baryon fraction assumed in this simulation. This alignment suggests a smaller depletion of baryons due to feedback mechanisms in TNG300 compared to Magneticum.

    \item Behaviour of scaling relation parameters: In the Magneticum simulations, the recovered posteriors for the scaling relation parameters closely match the fiducial values. However, in the TNG simulations, we observe a notable shift toward lower values ---of approximately 1$\sigma$--- in the posteriors for $C_\lambda$, that is, the parameter that controls redshift evolution. This shift indicates that the scaling relation in TNG absorbs some of the baryonic impacts. Such absorption by $C_\lambda$ is somewhat expected because it is the sole parameter in our mass-observation relation that evolves with background evolution, similar to the baryon fraction. The specific redshift dependency of the baryon fraction as a function of mass for TNG300 is detailed in Table~\ref{tab:par}.
\end{enumerate}

We note that, in the ~\citet{SPT:2018njh} cluster cosmology analysis from the SPT-SZ Survey, $C_\lambda$ is especially degenerate with the dark energy equation of state parameter $w,$  which we have not considered as a cosmological parameter for our forecast. The absorption of the baryonic impact on $C_\lambda$ is likely to lead to biased dark energy constraints in this case. 

In Table~\ref{tab:IOI&FOM}, we also report the degradation in the constraints on cosmological parameters after also marginalizing over the limited knowledge of the baryon content and the model accuracy, which are described by the $\alpha$ and $\beta$ parameters (see Eqs.~\ref{eq:alpha} and~\ref{eq:beta}). We present the relative change in the figure of merit~\citep[FOM; see][]{Huterer:2000mj,Albrecht:2006um} in the $(\Omega_{\rm m},\sigma_8)$ plane for different scenarios concerning hypothetical perfect knowledge of the baryonic content and a perfect performance of the baryonic correction model. Again, we consider $1$ and $3$ percent priors on the richness-mass scaling relations. In all cases, we assume a Gaussian prior on $\beta$ with zero mean and $0.01$ standard deviation in agreement with the accuracy and precision of our model. Concerning the priors on $\alpha$, we consider three scenarios: $0$ (perfect knowledge), $1$ percent, and $10$ percent. To provide some context,~\citet{DES:2017pjh} claim that the characteristic baryon mass fraction inside $R_{500{\rm c}}$ is about ($12.8 \pm 1.3$) percent for clusters with a mass of $M_{500{\rm c}}=4.8\times10^{14}\,M_\odot$ and at redshift $z = 0.6$ based on their analysis of a set of $91$ galaxy clusters from the $2500$ deg$^2$ South Pole Telescope SPT-SZ survey~\citep{SPT:2014wbo}. Therefore, our value of $10$ percent for the prior on $\alpha$ is in line with the current level of precision in the observational calibration of this relation. For $1$ and $3$ percent priors in the scaling relation and perfect knowledge of the baryon fraction relation, we observe that marginalizing over the accuracy of our model impacts the FOM in the cosmological constraints by $15$ and $11$ percent, respectively. While $1$ percent prior in the baryon fraction relation does not significantly further degrade the FOM for the case where the richness--mass relation is known at $3$ percent, it reduces the FOM by a further $7$ percent if the richness--mass relation is known with $1$ percent precision. Lastly, for a $10$ percent prior in the baryon fraction relation, the impact in the FOM is roughly $70$ and $40$ percent for the cases where the richness--mass relation is known at $1$ and $3$ percent precision level{, respectively}.

For all cases considered for the forecast, the correlation between $\alpha$ and $\beta$ and the cosmological parameters is less significant than the correlation between the cosmological parameters themselves. Although the absence of strong correlations between cosmological parameters and $\alpha$ and $\beta$ is a desired feature in order to improve the robustness of the  analysis, in the specific case of $\alpha$, it is also due to the simplified functional form considered for the uncertainty on the baryon fraction relation. A more detailed analysis of the impact of the baryon fraction relation, considering a more flexible parametrisation, is out of the scope of this paper and is left for further investigation.

In summary, the results presented in this section confirm that neglecting baryonic effects on halo masses leads to significantly biased constraints on cosmological parameters from the cluster number counts, especially when strong priors on the mass--observable relation are assumed. On the other hand, we also demonstrate that our model, which is designed to correct for such baryonic effects, significantly reduces this bias.

\begin{table*}
    \centering
    \caption{Summary statistics for the forecast of the impact of the baryon mass correction model on \Euclid's cluster count cosmological constraints on $\Omega_{\rm m}$ and $\sigma_8$. The IOI quantifies the tension in the posteriors for the different treatments with respect to hypothetical perfect knowledge of the model used to create the synthetic data (see text). The FOM ratio assesses the attenuation of the constraining power of the  cluster counts once one marginalises the uncertainties of the model and the baryon content of the clusters. For both statistics, errors were estimated by sampling multivariate Gaussian distributions with the correlation extracted from the respective MCMC chains and recomputing the statistics from them.}
    \label{tab:IOI&FOM}
    {\renewcommand{\arraystretch}{1.15} %<- modify value to suit your needs
    \begin{tabular}{c c c c c c}
                            &  & &\multicolumn{2}{c}{$\langle f_{\rm b, vir} (M,z) \rangle$} &  \\\cline{4-5}
    Summary statistics      & richness--mass relation priors & baryonic treatment  & relation & prior on $\alpha$ & value  \\ \hline 
    \multirow{4}{*}{IOI}
                            & \multirow{2}{*}{$1\,\%$}  & \multirow{2}{*}{ignoring} & TNG300 & \multirow{2}{*}{--}  &$\hphantom{-}1.8 \pm 0.1$  \\
                            &                           &  & Magneticum  &   &$\hphantom{-}4.1 \pm 0.2$  \\\cline{2-6}
                            & \multirow{2}{*}{$3\,\%$}& \multirow{2}{*}{ignoring} & TNG300  & \multirow{2}{*}{--}  &$\hphantom{-}0.9 \pm 0.1$  \\
                            &                           &  & Magneticum  &   &$\hphantom{-}2.0 \pm 0.2$  \\
                            \hline \hline
    \multirow{6}{*}{\scalebox{1.2}{$\frac{\mathrm{FOM}_\mathrm{corrected}}{\mathrm{FOM}}$}}
                            & \multirow{3}{*}{$1\,\%$} & \multirow{3}{*}{our model} & \multirow{3}{*}{Magneticum} & perfect knowledge  &$\hphantom{-}0.85 \pm 0.02$  \\
                            &  &                          &                          & $1\,\%$  &$\hphantom{-}0.78 \pm 0.02$  \\
                            &  &                          &                          & $10\,\%$  &$\hphantom{-}0.33 \pm 0.01$  \\
                            %&  &                          &                          & $50\,\%$  &$\hphantom{-}0.10 \pm 0.01$  \\
                            \cline{2-6}
                            & \multirow{3}{*}{$3\,\%$} & \multirow{3}{*}{our model}  & \multirow{3}{*}{Magneticum} & perfect knowledge  &$\hphantom{-}0.89 \pm 0.03$  \\
                            &  &                          &                          & $1\,\%$  &$\hphantom{-}0.90 \pm 0.03$  \\
                            &  &                          &                          & $10\,\%$  &$\hphantom{-}0.61 \pm 0.02$  \\
                            %&  &                          &                          & $50\,\%$  &$\hphantom{-}0.18 \pm 0.01$  \\
                            \hline \hline
    \end{tabular}}
\end{table*}

\section{\label{sec:conclusions}Conclusions}

In this paper, we present a model of the impact of baryonic feedback on the virial masses of galaxy clusters. The main aim of our analysis is to verify whether such baryonic effects on halo masses can be modelled with sufficient precision that they do not represent a limiting factor in the cosmological constraining power of the survey of galaxy clusters that will be obtained during the \Euclid wide photometric survey. 

Our model assumes that the feedback effects on the intracluster gas show a quasi-adiabatic behaviour and have the baryon fraction inside the virial radius as input. In this case, this effect can be reliably described by a phenomenological model that depends only on two free parameters and can be calibrated against cosmological hydro simulations. These two parameters control the minimum baryonic depletion observed in non-radiative simulations and the deviation from the adiabatic behaviour. Although our model is calibrated using a set of non-radiative hydrodynamic simulations and a single realisation of a full-physics simulation, we demonstrate that its performance is resilient to changes in the subresolution physics and cosmological parameters. This result is established by confronting our model with different simulations from the Magneticum suite and the TNG300 simulation. 
Finally, with the resulting baryonic correction model, we assess the impact of its accuracy and precision in the cosmological parameters inference from an idealised \Euclid cluster number count experiment. Our main conclusions can be summarised as follows.
\begin{itemize}
\item We show that the baryonic feedback effects on the intracluster gas can be accurately modelled using a quasi-adiabatic approach with two parameters: one controlling the minimum baryonic depletion and the other the deviation from adiabatic behaviour.
\item We successfully calibrated these two parameters against cosmological hydro simulations. This model calibration was conducted using a set of non-radiative hydrodynamic simulations as well as a single full-physics simulation.
\item Our model attains one percent relative accuracy in determining the virial dmo equivalent mass of simulated galaxy clusters.
\item The robustness of our model is demonstrated using different simulations from the Magneticum suite and the TNG300 simulation. Our findings indicate that the model's effectiveness does not significantly deviate with changes to the subresolution physics or cosmological parameters, showing strong consistency between our model predictions and these independent simulations.
\item Unlike previous works~\citep{Cui:2014aga,Bocquet:2015pva,Castro:2020yes}, our model does not rely on a single baryonic physics model or assume the universality of the HMF as a function of cosmology. This flexibility represents a significant advantage over other calibrations of the impact of baryonic feedback on the HMF, which cannot predict outcomes under altered baryonic models and implicitly assume HMF universality.
\item The findings of our research substantiate previous claims concerning the potentially significant impact of baryons ---if neglected--- on the cosmological constraints derived from the \Euclid photometric galaxy cluster number counts.
\item Most importantly, the uncertainties linked to our model for correcting baryonic effects on cluster masses are shown to be subdominant to both the precision of the expected calibration of the mass--observable relation with \Euclid and our current understanding of the baryon fraction within galaxy clusters.
\end{itemize}

Lastly, it is important to note that the model presented in this paper is based on the assumption that AGN feedback acts in a quasi-adiabatic manner. Future work could extend the current framework by investigating its validity concerning other prescriptions for AGN feedback as well as additional non-thermal processes not explored in this suite of simulations.

% Add the acknowledgement using the achnowledgements environment.
% Do not use \acknowledgement{\dots} as this affects the formatting
% of the references.
%

\begin{acknowledgements}
  
TC, SB, AS and AF are supported by the INFN INDARK PD51 grant. TC, SB and AS are supported by the Fondazione ICSC National Recovery and Resilience Plan (PNRR)
Project ID CN-00000013 ``Italian Research Center on High-Performance Computing, Big Data and Quantum Computing'' funded by MUR Missione 4 Componente 2 Investimento 1.4: ``Potenziamento strutture di ricerca e creazione di ``campioni nazionali'' di R$\&$S (M4C2-19)'' -- Next Generation EU (NGEU). TC and AS are also supported by the FARE MIUR grant `ClustersXEuclid' R165SBKTMA. AS is also supported by the ERC `ClustersXCosmo' grant agreement 716762. KD acknowledges support by the Deutsche Forschungsgemeinschaft (DFG, German Research Foundation) under Germany's Excellence Strategy -- EXC-2094 -- 390783311 as well as support through the COMPLEX project from the European Research Council (ERC) under the European Union's Horizon 2020 research and innovation program grant agreement ERC-2019-AdG 882679. AR is supported by the PRIN-MIUR 2017 WSCC32 ZOOMING grant. We acknowledge the computing centre of CINECA and INAF, under the coordination of the ``Accordo Quadro (MoU) per lo svolgimento di attivit{\`a} congiunta di ricerca Nuove frontiere in Astrofisica: HPC e Data Exploration di nuova generazione'', for the availability of computing resources and support. AF acknowledges support from Brookhaven National Laboratory. We acknowledge the use of the HOTCAT computing infrastructure of the Astronomical Observatory of Trieste -- National Institute for Astrophysics (INAF, Italy) \citep[see,][]{2020ASPC..527..303B,2020ASPC..527..307T}.  \AckEC
  
\end{acknowledgements}

%
% Here comes the reference list, generated via bibtex from
% your bibfile temp.bib
%

\bibliography{euclid}

\begin{thebibliography}{110}
\expandafter\ifx\csname natexlab\endcsname\relax\def\natexlab#1{#1}\fi

\bibitem[{Abbott {et~al.}(2020)Abbott, Aguena, Alarcon, {et~al.}}]{DES:2020ahh}
Abbott, T. M.~C., Aguena, M., Alarcon, A., {et~al.} 2020, PRD, 102, 023509

\bibitem[{Albrecht {et~al.}(2006)Albrecht, Bernstein, Cahn,
  {et~al.}}]{Albrecht:2006um}
Albrecht, A., Bernstein, G., Cahn, R., {et~al.} 2006, arXiv:0609591

\bibitem[{Allen {et~al.}(2011)Allen, Evrard, \& Mantz}]{Allen:2011zs}
Allen, S.~W., Evrard, A.~E., \& Mantz, A.~B. 2011, Ann. Rev. A\&A, 49, 409

\bibitem[{Andreon {et~al.}(2017)Andreon, Wang, Trinchieri, Moretti, \&
  Serra}]{Andreon:2017rih}
Andreon, S., Wang, J., Trinchieri, G., Moretti, A., \& Serra, A.~L. 2017, A\&A,
  606, A24

\bibitem[{Aric\`o {et~al.}(2021)Aric\`o, Angulo, Contreras,
  {et~al.}}]{Arico:2020lhq}
Aric\`o, G., Angulo, R.~E., Contreras, S., {et~al.} 2021, MNRAS, 506, 4070

\bibitem[{Beck {et~al.}(2016)Beck, Murante, Arth, {et~al.}}]{Beck:2015qva}
Beck, A.~M., Murante, G., Arth, A., {et~al.} 2016, MNRAS, 455, 2110

\bibitem[{{Bertocco} {et~al.}(2020){Bertocco}, {Goz}, {Tornatore},
  {et~al.}}]{2020ASPC..527..303B}
{Bertocco}, S., {Goz}, D., {Tornatore}, L., {et~al.} 2020, in Astronomical
  Society of the Pacific Conference Series, Vol. 527, Astronomical Data
  Analysis Software and Systems XXIX, ed. R.~{Pizzo}, E.~R. {Deul}, J.~D.
  {Mol}, J.~{de Plaa}, \& H.~{Verkouter}, 303

\bibitem[{Bleem {et~al.}(2015)Bleem, Stalder, De~Haan, {et~al.}}]{SPT:2014wbo}
Bleem, L.~E., Stalder, B., De~Haan, T., {et~al.} 2015, ApJ Suppl., 216, 27

\bibitem[{Blumenthal {et~al.}(1986)Blumenthal, Faber, Flores,
  {et~al.}}]{Blumenthal:1985qy}
Blumenthal, G.~R., Faber, S.~M., Flores, R., {et~al.} 1986, ApJ, 301, 27

\bibitem[{Bocquet {et~al.}(2019)Bocquet, Dietrich, Schrabback,
  {et~al.}}]{SPT:2018njh}
Bocquet, S., Dietrich, J.~P., Schrabback, T., {et~al.} 2019, ApJ, 878, 55

\bibitem[{Bocquet {et~al.}(2020)Bocquet, Heitmann, Habib,
  {et~al.}}]{Bocquet:2020tes}
Bocquet, S., Heitmann, K., Habib, S., {et~al.} 2020, ApJ, 901, 5

\bibitem[{Bocquet {et~al.}(2016)Bocquet, Saro, Dolag,
  {et~al.}}]{Bocquet:2015pva}
Bocquet, S., Saro, A., Dolag, K., {et~al.} 2016, MNRAS, 456, 2361

\bibitem[{Bocquet {et~al.}(2015)Bocquet, Saro, Mohr, {et~al.}}]{SPT:2014wkb}
Bocquet, S., Saro, A., Mohr, J.~J., {et~al.} 2015, ApJ, 799, 214

\bibitem[{Bondi(1952)}]{Bondi:1952ni}
Bondi, H. 1952, MNRAS, 112, 195

\bibitem[{Bondi \& Hoyle(1944)}]{Bondi:1944jm}
Bondi, H. \& Hoyle, F. 1944, MNRAS, 104, 273

\bibitem[{Borgani \& Kravtsov(2011)}]{Borgani:2009cd}
Borgani, S. \& Kravtsov, A. 2011, Adv. Sci. Lett., 4, 204

\bibitem[{{Borgani} {et~al.}(2001){Borgani}, {Rosati}, {Tozzi},
  {et~al.}}]{borgani:2001}
{Borgani}, S., {Rosati}, P., {Tozzi}, P., {et~al.} 2001, \apj, 561, 13

\bibitem[{Bryan \& Norman(1998)}]{Bryan:1997dn}
Bryan, G.~L. \& Norman, M.~L. 1998, ApJ, 495, 80

\bibitem[{Bullock \& Boylan-Kolchin(2017)}]{Bullock:2017xww}
Bullock, J.~S. \& Boylan-Kolchin, M. 2017, Ann. Rev. A\&A, 55, 343

\bibitem[{Castignani \& Benoist(2016)}]{Castignani:2016lvp}
Castignani, G. \& Benoist, C. 2016, A\&A, 595, A111

\bibitem[{Castro {et~al.}(2020)Castro, Borgani, Dolag,
  {et~al.}}]{Castro:2020yes}
Castro, T., Borgani, S., Dolag, K., {et~al.} 2020, MNRAS, 500, 2316

\bibitem[{Castro {et~al.}(2018)Castro, Quartin, Giocoli,
  {et~al.}}]{Castro:2017tbn}
Castro, T., Quartin, M., Giocoli, C., {et~al.} 2018, MNRAS, 478, 1305

\bibitem[{Chiu {et~al.}(2018)Chiu, Mohr, McDonald, {et~al.}}]{DES:2017pjh}
Chiu, I., Mohr, J.~J., McDonald, M., {et~al.} 2018, MNRAS, 478, 3072

\bibitem[{Costanzi {et~al.}(2019)Costanzi, Rozo, Simet, {et~al.}}]{DES:2018crd}
Costanzi, M., Rozo, E., Simet, M., {et~al.} 2019, MNRAS, 488, 4779

\bibitem[{Costanzi {et~al.}(2021)Costanzi, Saro, Bocquet,
  {et~al.}}]{DES:2020cbm}
Costanzi, M., Saro, A., Bocquet, S., {et~al.} 2021, PRD, 103, 043522

\bibitem[{Crain {et~al.}(2015)Crain, Schaye, Bower, {et~al.}}]{Crain:2015poa}
Crain, R.~A., Schaye, J., Bower, R.~G., {et~al.} 2015, MNRAS, 450, 1937

\bibitem[{Cui {et~al.}(2014)Cui, Borgani, \& Murante}]{Cui:2014aga}
Cui, W., Borgani, S., \& Murante, G. 2014, MNRAS, 441, 1769

\bibitem[{Davis {et~al.}(1985)Davis, Efstathiou, Frenk,
  {et~al.}}]{Davis:1985rj}
Davis, M., Efstathiou, G., Frenk, C.~S., {et~al.} 1985, ApJ, 292, 371

\bibitem[{Debackere {et~al.}(2021)Debackere, Schaye, \&
  Hoekstra}]{Debackere:2021ado}
Debackere, S. N.~B., Schaye, J., \& Hoekstra, H. 2021, MNRAS, 505, 593

\bibitem[{Despali {et~al.}(2016)Despali, Giocoli, Angulo,
  {et~al.}}]{Despali:2015yla}
Despali, G., Giocoli, C., Angulo, R.~E., {et~al.} 2016, MNRAS, 456, 2486

\bibitem[{Di~Matteo {et~al.}(2008)Di~Matteo, Colberg, Springel,
  {et~al.}}]{DiMatteo:2007sq}
Di~Matteo, T., Colberg, J., Springel, V., {et~al.} 2008, ApJ, 676, 33

\bibitem[{Diemer \& Joyce(2019)}]{Diemer:2018vmz}
Diemer, B. \& Joyce, M. 2019, ApJ, 871, 168

\bibitem[{{Dolag} {et~al.}(2009){Dolag}, {Borgani}, {Murante},
  {et~al.}}]{Dolag:2009}
{Dolag}, K., {Borgani}, S., {Murante}, G., {et~al.} 2009, \mnras, 399, 497

\bibitem[{Duffy {et~al.}(2010)Duffy, Schaye, Kay, Dalla~Vecchia, Battye, \&
  Booth}]{Duffy:2010hf}
Duffy, A.~R., Schaye, J., Kay, S.~T., {et~al.} 2010, MNRAS, 405, 2161

\bibitem[{{Ellien} {et~al.}(2019){Ellien}, {Durret}, {Adami},
  {et~al.}}]{2019A&A...628A..34E}
{Ellien}, A., {Durret}, F., {Adami}, C., {et~al.} 2019, \aap, 628, A34

\bibitem[{Euclid Collaboration:~Adam {et~al.}(2019)Euclid Collaboration:~Adam,
  Vannier, Maurogordato, {et~al.}}]{Euclid:2019bue}
Euclid Collaboration:~Adam, R., Vannier, M., Maurogordato, S., {et~al.} 2019,
  A\&A, 627, A23

\bibitem[{Euclid Collaboration:~Castro {et~al.}(2023)Euclid
  Collaboration:~Castro, Fumagalli, Angulo, {et~al.}}]{Euclid:2022dbc}
Euclid Collaboration:~Castro, T., Fumagalli, F., Angulo, R.~E., {et~al.} 2023,
  A\&A, 671, A100

\bibitem[{Euclid Collaboration:~Deshpande {et~al.}(2023)Euclid
  Collaboration:~Deshpande, Kitching, Hall, {et~al.}}]{Euclid:2023wdq}
Euclid Collaboration:~Deshpande, A.~C., Kitching, T., Hall, A., {et~al.} 2023,
  arXiv:2302.04507

\bibitem[{Euclid Collaboration:~Fumagalli {et~al.}(2021)Euclid
  Collaboration:~Fumagalli, Saro, Borgani, {et~al.}}]{Euclid:2021api}
Euclid Collaboration:~Fumagalli, A., Saro, A., Borgani, S., {et~al.} 2021,
  A\&A, 652, A21

\bibitem[{Euclid Collaboration:~Scaramella {et~al.}(2022)Euclid
  Collaboration:~Scaramella, Amiaux, Mellier, {et~al.}}]{2022A&A...662A.112E}
Euclid Collaboration:~Scaramella, R., Amiaux, J., Mellier, Y., {et~al.} 2022,
  A\&A, 662, A112

\bibitem[{Ferland {et~al.}(1998)Ferland, Korista, Verner,
  {et~al.}}]{Ferland:1998id}
Ferland, G.~J., Korista, K.~T., Verner, D.~A., {et~al.} 1998, Publ. Astron.
  Soc. Pac., 110, 761

\bibitem[{Fumagalli {et~al.}(2023)Fumagalli, Costanzi, Saro,
  {et~al.}}]{Fumagalli:2023yym}
Fumagalli, A., Costanzi, M., Saro, A., {et~al.} 2023, arXiv:2310.09146

\bibitem[{Gnedin {et~al.}(2004)Gnedin, Kravtsov, Klypin,
  {et~al.}}]{Gnedin:2004cx}
Gnedin, O.~Y., Kravtsov, A.~V., Klypin, A.~A., {et~al.} 2004, ApJ, 616, 16

\bibitem[{Hasselfield {et~al.}(2013)Hasselfield, Hilton, Marriage,
  {et~al.}}]{Hasselfield:2013wf}
Hasselfield, M., Hilton, M., Marriage, T.~A., {et~al.} 2013, JCAP, 07, 008

\bibitem[{Hirschmann {et~al.}(2014)Hirschmann, Dolag, Saro,
  {et~al.}}]{Hirschmann:2013qfl}
Hirschmann, M., Dolag, K., Saro, A., {et~al.} 2014, MNRAS, 442, 2304

\bibitem[{Holder {et~al.}(2001)Holder, Haiman, \& Mohr}]{holder:2001db}
Holder, G., Haiman, Z., \& Mohr, J. 2001, ApJ Lett., 560, L111

\bibitem[{Hoyle \& Lyttleton(1939)}]{hoyle_lyttleton_1939}
Hoyle, F. \& Lyttleton, R.~A. 1939, Mathematical Proceedings of the Cambridge
  Philosophical Society, 35, 405

\bibitem[{Hu \& Kravtsov(2003)}]{Hu:2002we}
Hu, W. \& Kravtsov, A.~V. 2003, ApJ, 584, 702

\bibitem[{Huterer \& Turner(2001)}]{Huterer:2000mj}
Huterer, D. \& Turner, M.~S. 2001, PRD, 64, 123527

\bibitem[{Jesseit {et~al.}(2002)Jesseit, Naab, \& Burkert}]{Jesseit:2002tj}
Jesseit, R., Naab, T., \& Burkert, A. 2002, ApJ Lett., 571, L89

\bibitem[{Karamanis \& Beutler(2021)}]{karamanis2020ensemble}
Karamanis, M. \& Beutler, F. 2021, Stat. Comput., 31, 61

\bibitem[{Karamanis {et~al.}(2021)Karamanis, Beutler, \&
  Peacock}]{karamanis2021zeus}
Karamanis, M., Beutler, F., \& Peacock, J.~A. 2021, MNRAS, 508, 3589

\bibitem[{Komatsu {et~al.}(2011)Komatsu, Dunkley, Nolta,
  {et~al.}}]{WMAP:2010qai}
Komatsu, E., Dunkley, J., Nolta, M.~R., {et~al.} 2011, ApJ Suppl., 192, 18

\bibitem[{Kravtsov \& Borgani(2012)}]{Kravtsov:2012zs}
Kravtsov, A. \& Borgani, S. 2012, Ann. Rev. A\&A, 50, 353

\bibitem[{{Laureijs} {et~al.}(2011){Laureijs}, {Amiaux}, {Arduini},
  {et~al.}}]{2011arXiv1110.3193L}
{Laureijs}, R., {Amiaux}, J., {Arduini}, S., {et~al.} 2011, arXiv:1110.3193

\bibitem[{Lesci {et~al.}(2022)Lesci, Marulli, F., Moscardini, L.,
  {et~al.}}]{Lesci:2020qpk}
Lesci, G.~F., Marulli, F., {et~al.} 2022, A\&A, 659, A88

\bibitem[{Lin \& Ishak(2017)}]{Lin:2017ikq}
Lin, W. \& Ishak, M. 2017, PRD, 96, 023532

\bibitem[{Mantz {et~al.}(2015)Mantz, Von~der Linden, Allen,
  {et~al.}}]{Mantz:2014paa}
Mantz, A.~B., Von~der Linden, A., Allen, S.~W., {et~al.} 2015, MNRAS, 446, 2205

\bibitem[{Martizzi {et~al.}(2013)Martizzi, T., \& Moore}]{Martizzi:2012ci}
Martizzi, D., T., R., \& Moore, B. 2013, MNRAS, 432, 1947

\bibitem[{McCarthy {et~al.}(2017)McCarthy, Schaye, Bird,
  {et~al.}}]{McCarthy:2016mry}
McCarthy, I.~G., Schaye, J., Bird, S., {et~al.} 2017, MNRAS, 465, 2936

\bibitem[{McDonald {et~al.}(2012)McDonald, Bayliss, Benson,
  {et~al.}}]{McDonald:2012hz}
McDonald, M., Bayliss, M., Benson, A., {et~al.} 2012, Nature, 488, 349

\bibitem[{{Naiman} {et~al.}(2018){Naiman}, {Pillepich}, {Springel},
  {Ramirez-Ruiz}, {Torrey}, {Vogelsberger}, {Pakmor}, {Nelson}, {Marinacci},
  {Hernquist}, {Weinberger}, \& {Genel}}]{2018MNRAS.477.1206N}
{Naiman}, J.~P., {Pillepich}, A., {Springel}, V., {et~al.} 2018, MNRAS, 477,
  1206

\bibitem[{Navarro {et~al.}(1997)Navarro, Frenk, \& White}]{Navarro:1996gj}
Navarro, J.~F., Frenk, C.~S., \& White, S. D.~M. 1997, ApJ, 490, 493

\bibitem[{{Nelson} {et~al.}(2019){Nelson}, {Springel}, {Pillepich},
  {Rodriguez-Gomez}, {Torrey}, {Genel}, {Vogelsberger}, {Pakmor}, {Marinacci},
  {Weinberger}, {Kelley}, {Lovell}, {Diemer}, \&
  {Hernquist}}]{2019ComAC...6....2N}
{Nelson}, D., {Springel}, V., {Pillepich}, A., {et~al.} 2019, Computational
  Astrophysics and Cosmology, 6, 2

\bibitem[{Nelson {et~al.}(2018)}]{Nelson:2017cxy}
Nelson, D. {et~al.} 2018, MNRAS, 475, 624

\bibitem[{Pakmor {et~al.}(2011)Pakmor, Bauer, \& Springel}]{Pakmor:2011ht}
Pakmor, R., Bauer, A., \& Springel, V. 2011, MNRAS, 418, 1392

\bibitem[{Pakmor {et~al.}(2014)Pakmor, Marinacci, \& Springel}]{Pakmor:2013rqa}
Pakmor, R., Marinacci, F., \& Springel, V. 2014, ApJ Lett., 783, L20

\bibitem[{Pakmor \& Springel(2013)}]{Pakmor:2012xy}
Pakmor, R. \& Springel, V. 2013, MNRAS, 432, 176

\bibitem[{Paranjape {et~al.}(2021)Paranjape, Choudhury, \&
  Sheth}]{Paranjape:2021zia}
Paranjape, A., Choudhury, T.~R., \& Sheth, R.~K. 2021, MNRAS, 503, 4147

\bibitem[{Pillepich {et~al.}(2018{\natexlab{a}})Pillepich, Springel, Nelson,
  {et~al.}}]{Pillepich:2017jle}
Pillepich, A., Springel, V., Nelson, D., {et~al.} 2018{\natexlab{a}}, MNRAS,
  473, 4077

\bibitem[{Pillepich {et~al.}(2018{\natexlab{b}})}]{Pillepich:2017fcc}
Pillepich, A. {et~al.} 2018{\natexlab{b}}, MNRAS, 475, 648

\bibitem[{{Planck Collaboration XIII.}(2016)}]{Planck:2015fie}
{Planck Collaboration XIII.} 2016, A\&A, 594, A13

\bibitem[{{Planck Collaboration XX.}(2014)}]{Planck:2013lkt}
{Planck Collaboration XX.} 2014, A\&A, 571, A20

\bibitem[{{Planck Collaboration XXIV.}(2016)}]{Planck:2015lwi}
{Planck Collaboration XXIV.} 2016, A\&A, 594, A24

\bibitem[{Rozo {et~al.}(2010)Rozo, Wechsler, Rykoff, {et~al.}}]{DSDD:2009php}
Rozo, E., Wechsler, R.~H., Rykoff, E.~S., {et~al.} 2010, ApJ, 708, 645

\bibitem[{{Ryden} \& {Gunn}(1987)}]{1987ApJ...318...15R}
{Ryden}, B.~S. \& {Gunn}, J.~E. 1987, \apj, 318, 15

\bibitem[{Saro {et~al.}(2015)Saro, Bocquet, Rozo, {et~al.}}]{DES:2015mqu}
Saro, A., Bocquet, S., Rozo, E., {et~al.} 2015, MNRAS, 454, 2305

\bibitem[{Sartoris {et~al.}(2016)Sartoris, Biviano, Fedeli,
  {et~al.}}]{Sartoris:2015aga}
Sartoris, B., Biviano, A., Fedeli, C., {et~al.} 2016, MNRAS, 459, 1764

\bibitem[{Sawala {et~al.}(2016)Sawala, Frenk, Fattahi,
  {et~al.}}]{Sawala:2015cdf}
Sawala, T., Frenk, C.~S., Fattahi, A., {et~al.} 2016, MNRAS, 457, 1931

\bibitem[{Schaye {et~al.}(2015)Schaye, Crain, Bower, {et~al.}}]{Schaye:2014tpa}
Schaye, J., Crain, R.~A., Bower, R.~G., {et~al.} 2015, MNRAS, 446, 521

\bibitem[{Schaye {et~al.}(2023)Schaye, Kugel, Schaller, Helly, Braspenning,
  Elbers, McCarthy, van Daalen, Vandenbroucke, Frenk, Kwan, Salcido, Bahé,
  Borrow, Chaikin, Hahn, Huško, Jenkins, Lacey, \&
  Nobels}]{10.1093/mnras/stad2419}
Schaye, J., Kugel, R., Schaller, M., {et~al.} 2023, MNRAS, stad2419

\bibitem[{{Schellenberger} {et~al.}(2019){Schellenberger}, {David},
  {O'Sullivan}, {Vrtilek}, \& {Haines}}]{Schellenberger:2019aws}
{Schellenberger}, G., {David}, L., {O'Sullivan}, E., {Vrtilek}, J.~M., \&
  {Haines}, C.~P. 2019, \apj, 882, 59

\bibitem[{Schneider \& Teyssier(2015)}]{Schneider:2015wta}
Schneider, A. \& Teyssier, R. 2015, JCAP, 12, 049

\bibitem[{Singh {et~al.}(2020)Singh, Saro, Costanzi, {et~al.}}]{Singh:2019end}
Singh, P., Saro, A., Costanzi, M., {et~al.} 2020, MNRAS, 494, 3728

\bibitem[{Springel(2005)}]{Springel:2005mi}
Springel, V. 2005, MNRAS, 364, 1105

\bibitem[{Springel(2010)}]{Springel:2009aa}
Springel, V. 2010, MNRAS, 401, 791

\bibitem[{Springel {et~al.}(2005)Springel, Di~Matteo, \&
  Hernquist}]{Springel:2004kf}
Springel, V., Di~Matteo, T., \& Hernquist, L. 2005, MNRAS, 361, 776

\bibitem[{{Springel} \& {Hernquist}(2003)}]{Springel:2003}
{Springel}, V. \& {Hernquist}, L. 2003, \mnras, 339, 289

\bibitem[{Springel {et~al.}(2021)Springel, Pakmor, Zier,
  {et~al.}}]{Springel:2020plp}
Springel, V., Pakmor, R., Zier, O., {et~al.} 2021, MNRAS, 506, 2871

\bibitem[{Springel {et~al.}(2001{\natexlab{a}})Springel, White, Tormen,
  {et~al.}}]{Springel:2000qu}
Springel, V., White, S. D.~M., Tormen, G., {et~al.} 2001{\natexlab{a}}, MNRAS,
  328, 726

\bibitem[{Springel {et~al.}(2001{\natexlab{b}})Springel, Yoshida, \&
  White}]{Springel:2000yr}
Springel, V., Yoshida, N., \& White, S. D.~M. 2001{\natexlab{b}}, New Astron.,
  6, 79

\bibitem[{Springel {et~al.}(2018)}]{Springel:2017tpz}
Springel, V. {et~al.} 2018, MNRAS, 475, 676

\bibitem[{{Steigman} {et~al.}(1978){Steigman}, {Sarazin}, {Quintana},
  {et~al.}}]{1978AJ.....83.1050S}
{Steigman}, G., {Sarazin}, C.~L., {Quintana}, H., {et~al.} 1978, \aj, 83, 1050

\bibitem[{{Taffoni} {et~al.}(2020){Taffoni}, {Becciani}, {Garilli},
  {et~al.}}]{2020ASPC..527..307T}
{Taffoni}, G., {Becciani}, U., {Garilli}, B., {et~al.} 2020, in Astronomical
  Society of the Pacific Conference Series, Vol. 527, Astronomical Data
  Analysis Software and Systems XXIX, ed. R.~{Pizzo}, E.~R. {Deul}, J.~D.
  {Mol}, J.~{de Plaa}, \& H.~{Verkouter}, 307

\bibitem[{Teyssier {et~al.}(2011)Teyssier, Moore, Martizzi,
  {et~al.}}]{Teyssier:2010dp}
Teyssier, R., Moore, B., Martizzi, D., {et~al.} 2011, MNRAS, 414, 195

\bibitem[{Tinker {et~al.}(2008)Tinker, Kravtsov, Klypin,
  {et~al.}}]{Tinker:2008ff}
Tinker, J.~L., Kravtsov, A.~V., Klypin, A., {et~al.} 2008, ApJ, 688, 709

\bibitem[{Tornatore {et~al.}(2007)Tornatore, Borgani, Dolag,
  {et~al.}}]{Tornatore:2007ds}
Tornatore, L., Borgani, S., Dolag, K., {et~al.} 2007, MNRAS, 382, 1050

\bibitem[{van Daalen {et~al.}(2020)van Daalen, McCarthy, \&
  Schaye}]{vanDaalen:2019pst}
van Daalen, M.~P., McCarthy, I.~G., \& Schaye, J. 2020, MNRAS, 491, 2424

\bibitem[{van Daalen {et~al.}(2011)van Daalen, Schaye, Booth, \&
  Vecchia}]{vanDaalen:2011xb}
van Daalen, M.~P., Schaye, J., Booth, C.~M., \& Vecchia, C.~D. 2011, MNRAS,
  415, 3649

\bibitem[{Velliscig {et~al.}(2014)Velliscig, van Daalen, Schaye,
  {et~al.}}]{Velliscig:2014bza}
Velliscig, M., van Daalen, M.~P., Schaye, J., {et~al.} 2014, MNRAS, 442, 2641

\bibitem[{Velmani \& Paranjape(2023)}]{Velmani:2022una}
Velmani, P. \& Paranjape, A. 2023, MNRAS, 520, 2867

\bibitem[{Vogelsberger {et~al.}(2013)Vogelsberger, Genel, Sijacki,
  {et~al.}}]{Vogelsberger:2013eka}
Vogelsberger, M., Genel, S., Sijacki, D., {et~al.} 2013, MNRAS, 436, 3031

\bibitem[{Vogelsberger {et~al.}(2014)Vogelsberger, Genel, Springel,
  {et~al.}}]{Vogelsberger:2014dza}
Vogelsberger, M., Genel, S., Springel, V., {et~al.} 2014, MNRAS, 444, 1518

\bibitem[{Vogelsberger {et~al.}(2020)Vogelsberger, Marinacci, Torrey,
  {et~al.}}]{Vogelsberger:2019ynw}
Vogelsberger, M., Marinacci, F., Torrey, P., {et~al.} 2020, Nature Rev. Phys.,
  2, 42

\bibitem[{Watson {et~al.}(2013)Watson, Iliev, D'Aloisio,
  {et~al.}}]{Watson:2012mt}
Watson, W.~A., Iliev, I.~T., D'Aloisio, A., {et~al.} 2013, MNRAS, 433, 1230

\bibitem[{Webb {et~al.}(2015)Webb, Noble, DeGroot, {et~al.}}]{Webb:2015tha}
Webb, T., Noble, A., DeGroot, A., {et~al.} 2015, ApJ, 809, 173

\bibitem[{{Weinberger} {et~al.}(2017){Weinberger}, {Springel}, {Hernquist},
  {et~al.}}]{2017MNRAS.465.3291W}
{Weinberger}, R., {Springel}, V., {Hernquist}, L., {et~al.} 2017, \mnras, 465,
  3291

\bibitem[{Wiersma {et~al.}(2009)Wiersma, Schaye, \& Smith}]{Wiersma:2008cs}
Wiersma, R. P.~C., Schaye, J., \& Smith, B.~D. 2009, MNRAS, 393, 99

\bibitem[{{Yuan} {et~al.}(2020){Yuan}, {Elagali}, {Labb{\'e}},
  {et~al.}}]{2020NatAs...4..957Y}
{Yuan}, T., {Elagali}, A., {Labb{\'e}}, I., {et~al.} 2020, Nature Astronomy, 4,
  957

\bibitem[{{Zeldovich} {et~al.}(1980){Zeldovich}, {Klypin}, {Khlopov},
  {et~al.}}]{1980SvJNP..31..664Z}
{Zeldovich}, Y.~B., {Klypin}, A., {Khlopov}, M.~Y., {et~al.} 1980, Soviet
  Journal of Nuclear Physics, 31, 664

\end{thebibliography}

%
% Now you can add appendices.
% In this example, the appendices are in one column mode.
% If that is not requires, comment out \onecolumn
% Note that appendices in A\&A come {\it after\/} the references.

\begin{appendix}
%\onecolumn %If you don't want single column for the Appendix, please %comment this out
\section{Baryon content of galaxy clusters in full-physics simulations}\label{app:content}

In this Appendix, we report results pertaining to the baryon content of galaxy clusters and groups within the halo virial radius for both the Magneticum and the TNG300 simulations. The results provided here are used in our analysis for the calibration of the parameters that define our model to correct halo masses for baryonic effects. A comparison with observational data is beyond the scope of the present study.

\begin{figure*}
\centering
\includegraphics[width=\textwidth]{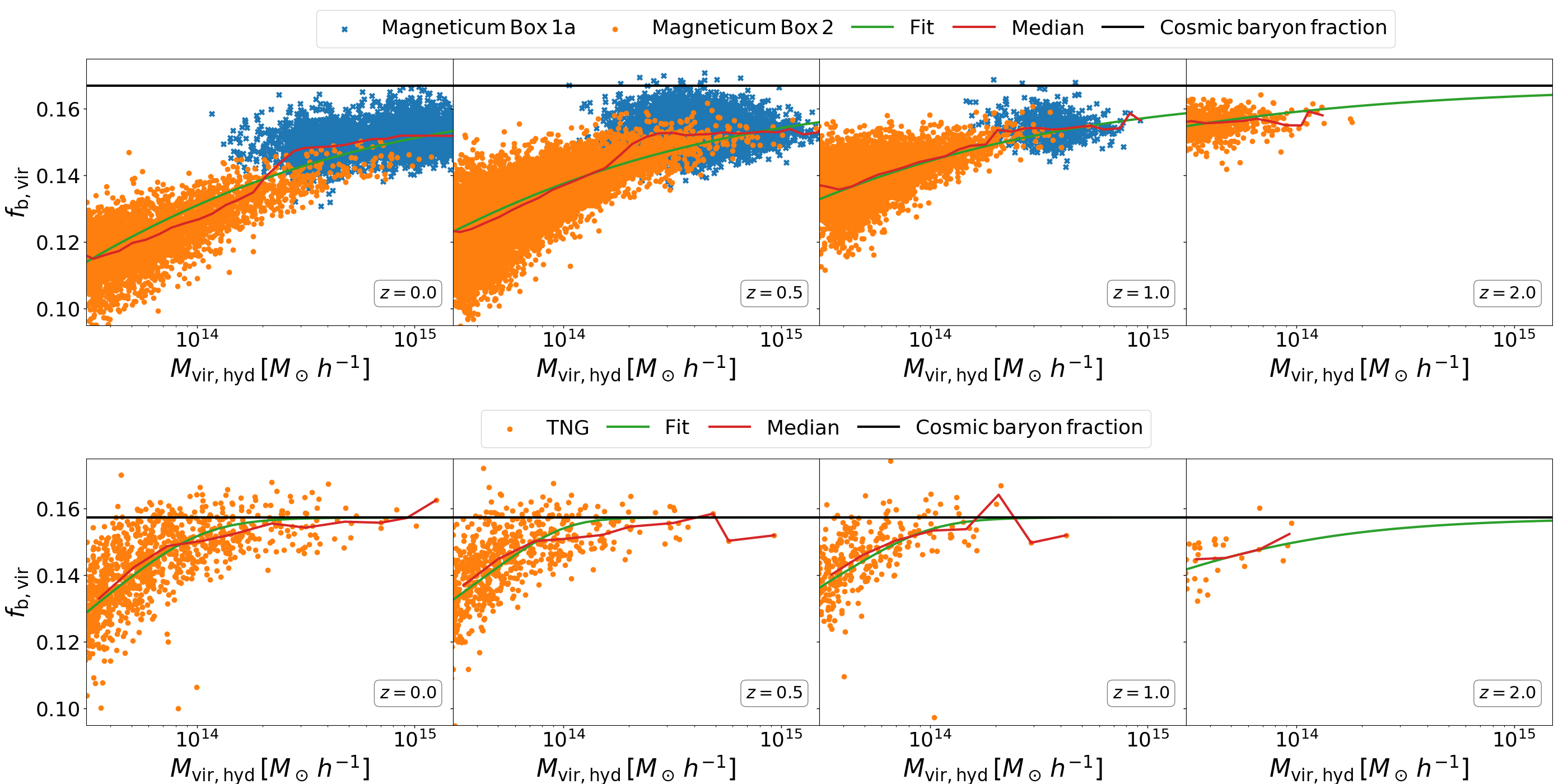}
\caption{Baryon fraction inside the virial radius as a function of the virial mass. \emph{Top:}  Magneticum suite of simulations. Orange and blue symbols correspond to Box 2 and Box 1a, respectively. \emph{Bottom:}  TNG300 simulation. Different columns correspond to the redshifts $z\in\{0.0, 0.5, 1.0, 2.0\}$. The assumed cosmic baryon fraction for each simulation is depicted as the horizontal black line.}
\label{fig:fb-sims}
\end{figure*}

In Fig.~\ref{fig:fb-sims} we show the baryonic fraction inside the virial halos as a function of the halo virial mass for both the Magneticum (upper panels) and TNG300 (lower panels) simulations, at four different redshifts $z\in\{0.0, 0.5, 1.0, 2.0\}$. The cosmic baryon fraction assumed by the respective simulations is shown as the horizontal black line. The median baryon fraction in halos is shown by the red line and the best fit for the relation is shown in green. For the Magneticum results, we show the results obtained for Box 2 and Box 1a with orange and blue symbols, respectively. Box 2 samples the mass range of groups and low-mass clusters  at
increased resolution, while the larger size of Box 1a allows us to sample the most massive clusters 
at lower resolution (see Table \ref{tab:magnorig}). 

The best-fit relations for the Magneticum and TNG300 are given by
\begin{equation}
    f_{\rm b, fid} (m, z) = f_{\rm b, cosmic} \, m^{-\gamma} \,\left( 1 + m^{1/\delta}  \right)^{\gamma\,\delta}\,,
    \label{eq:fb-sims}
\end{equation}
with $m$ given by
\begin{equation}
    m=\log_{10}\,\frac{M_{\rm vir}}{10^{14}\,M_\odot\,h^{-1}}\,,
\end{equation}
while the specific values for the parameters $\gamma$ and $\delta$ and their redshift dependence are shown in Table~\ref{tab:par}.

Interestingly, the depletion of baryons in the Magneticum simulations is more pronounced than in TNG300, a difference that increases towards lower redshifts. At $z=0$, a depletion of about 10 percent is observed for Magneticum even for the most massive clusters, while the baryon content in TNG300 clusters saturates to the cosmic value already for $M_{\rm vir}\simeq 2\times 10^{14} M_\odot\, h^{-1}$.
\begin{table}
    \centering
    \caption{Best-fit parameters for the fitting function presented in Eq.~\eqref{eq:fb-sims} for the baryon fraction inside the virial radius for the Magneticum and TNG300 simulations.}
    \resizebox{1.\columnwidth}{!}{%
    {\renewcommand{\arraystretch}{1.} %<- modify value to suit your needs
    \begin{tabular}{c|c|c}
    Simulation & $\gamma(z)$ & $\delta(z)$ \\\hline
    TNG300 & $0.4106 \, z -1.4640$ & $0.0397\,z^2 - 0.04358 \, z + 0.03567$ \\
    Magneticum &  $0.7008\,z -1.7505$ & $0.2$ \\\hline\hline
    \end{tabular}}}
    \label{tab:par}
\end{table}

\end{appendix}

\end{document}